\documentclass[reprint,amsmath,amssymb,aps]{revtex4-1}

\usepackage{graphicx} 
\usepackage{dcolumn}   
\usepackage{bm}        
\usepackage{amssymb}   
\usepackage{mathtools}
\usepackage{mathpazo}
\usepackage{lipsum}
\usepackage{hyperref}
\hypersetup{colorlinks,allcolors=black}
\usepackage{amsthm}
\usepackage{ragged2e}
\usepackage{verbatim}
\usepackage{color,xcolor,colortbl}
\usepackage{soul}
\usepackage{multirow}
\usepackage{boxedminipage}
\usepackage{changepage}
\usepackage{amsmath}
\usepackage{arydshln}

\newcommand{\Tr}{\text{Tr}}

\begin{document}
\preprint{APS/123-QED}

\title{Improving key rates of the unbalanced phase-encoded BB84 protocol using the flag-state squashing model}

\author{Nicky Kai Hong Li}
\author{Norbert L\"utkenhaus}
\affiliation{Institute for Quantum Computing and Department of Physics and Astronomy, University of Waterloo, Waterloo, Ontario, Canada N2L 3G1
}
\date{\today}

\begin{abstract}
	All phase-encoded BB84 implementations have signal states with unbalanced amplitudes in practice. Thus, the original security analyses a priori do not apply to them. Previous security proofs use signal tagging of multi-photon pulses to recover the behaviour of regular BB84. This is overly conservative, as for unbalanced signals, the photon-number splitting attack does not leak full information to Eve. In this work, we exploit the flag-state squashing model to preserve some parts of the multi-photon generated private information in our analysis. Using a numerical proof technique, we obtain significantly higher key rates compared with previously published results in the low-loss regime. It turns out that the usual scenario of untrusted dark counts runs into conceptual difficulties in some parameter regime. Thus, we discuss the trusted dark count scenario in this paper as well. We also report a gain in key rates when part of the total loss is known to be induced by a trusted device. We highlight that all these key rate improvements can be achieved without modification of the experimental setup.

\end{abstract}

\maketitle

\section{Introduction}

The earliest phase-encoding quantum key distribution (QKD) scheme was proposed by Bennett \cite{Bennett} in 1992 as a demonstration that any two non-orthogonal states can be used for generating shared secret keys between two parties. Later, Townsend \cite{Townsend1994} and then Hughes et al.\;\cite{Hughes} proposed a more practical phase-encoding BB84 protocol which uses two Mach-Zehnder interferometers. In practice, the phase-modulator in each Mach-Zehnder unit will introduce photon loss, thereby causing an asymmetry between the intensities of the phase-encoded pulse and the reference pulse even if the typical observations do not directly reveal this. This asymmetric loss was addressed in Refs. \cite{Li2009security, Agnes, Kiyoshi} which model the loss caused by an imperfect phase-modulator with a beam splitter (BS) of the same transmission probability.

The first attempt in giving security proofs for this protocol was made by Ref. \cite{Li2009security}. Formal security proofs were later on provided by Refs. \cite{Agnes,Kiyoshi} which both used qubit-based reduction proof techniques. Despite being a deviation from the standard BB84 protocol, Ref. \cite{Kiyoshi} confirms that the old security analysis for the balanced protocol still holds in the unbalanced case. This calls for a revision of the security statement made by Ref. \cite{Agnes}, which we will discuss in detail in Sec. \ref{ResultsSec}.

Both Refs. \cite{Agnes} and \cite{Kiyoshi} use decoy states \cite{Decoy0,Decoy1,Decoy2}, signal tagging \cite{Tagging1,Tagging2}, and the qubit squashing model \cite{Tagging1,SquashingPaper1,SquashingPaper2,VarunThesis} to convert the full security analysis into an effective qubit-to-qubit security analysis problem. Due to the asymmetric intensities of the signal states, the photon number splitting (PNS) attack \cite{PNSattack} will not leak full information of the signal's multi-photon part to Eve since in this case, a single photon obtained in the PNS attack will be in one of two non-orthogonal states, even after basis announcements. Thus, the tagging approach, which pessimistically assumes that all multi-photon signals leak their full information to an adversary, simplifies the security proof but underestimates the secure key rate of this protocol.

In this paper, we will answer the following questions: Could we improve the key rates in Ref. \cite{Kiyoshi} if we keep the multi-photon part of the signals? Could the multi-photon part of the signal contribute significantly to key rates when the total loss or the asymmetry is large?

To highlight the differences between our approach and Refs. \cite{Agnes,Kiyoshi}'s, we apply the numerical analysis formulated in \cite{AdamNumerics} which involves optimisations over finite-dimensional matrices to obtain reliable lower bounds on the key rates. On the source side, we treat lower photon numbers explicitly, while turning to tagging again for higher photon numbers. On the receiver side, we know that the qubit squashing model converts the multi-click events caused by the multi-photon part of the signals into additional qubit errors \cite{SquashingPaper2,Agnes,VarunThesis}. The convenience of reaching a qubit picture may thus cost a reduction in key rate. Therefore, we use the flag-state squashing model \cite{Yanbao} to circumvent this problem, especially for low-loss channels. The flag-state squashing model preserves any measurement on a low photon-number subspace, while tagging the arriving signals of higher photon numbers. As a result, we obtain secret key rates that can exceed the ones quoted in Refs. \cite{Agnes,Kiyoshi}.

During our investigations, we noticed a problem with the common approach which attributes all observed errors to an adversary and describes Bob's detection device by an idealised set-up. Once the actual detectors have some dark count rate, this approach may lead in some circumstances to unphysical constraints, meaning that such an ideal device could not lead to the actual observations. For that reason, we will also introduce results for trusted detector noises, especially dark counts, for which this problem does not exist.

The rest of this paper is outlined as follows. We first revisit the protocol in Sec. \ref{ProtocolDescription} and describe the mathematical model of the protocol in Sec. \ref{MathModel}. We will then justify our security proof techniques and state the methods that allow us to speed up our key rate computations
in Sec. \ref{SecurityProofTech}. With the description of how we simulate experimental statistics in Sec. \ref{SimulateExperiment}, we present our lower bounds for the secure key rates of the protocol in Sec. \ref{ResultsSec}. A summary of our results is provided in Sec. \ref{Conclusion} to conclude this paper. Full justifications of the proof techniques mentioned in Sec. \ref{SecurityProofTech} are discussed in the Appendices.
	
\section{Protocol description}\label{ProtocolDescription}

\begin{figure}[h!]
    \centering
    \includegraphics[width=1.05\linewidth]{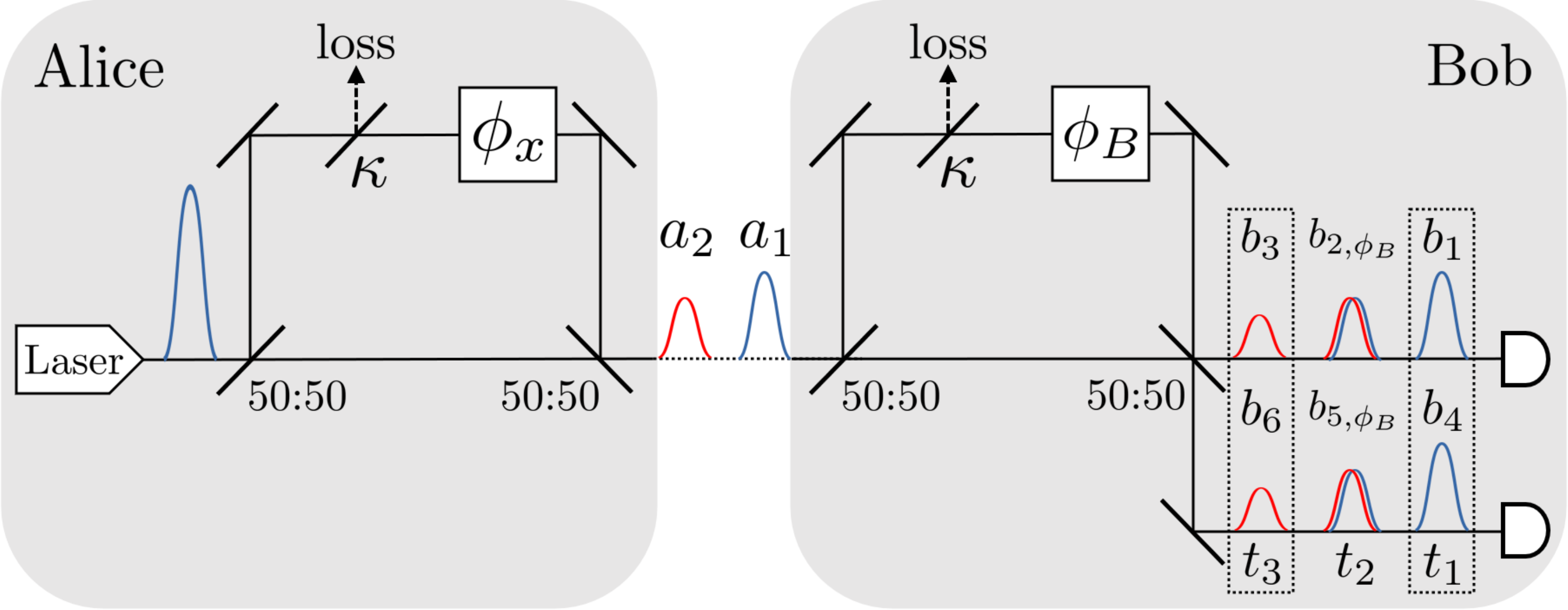}
    \caption{The setup for the unbalanced phase-encoded BB84 protocol. All beam splitters (BSs) are labelled by their transmissivities. The grouping of Bob's detection events are represented by the dotted boxes. \label{setup}}
\end{figure}

We consider a phase-encoded BB84 protocol with a Mach-Zehnder set-up. The only modification is that we take into account the typical loss in one arm of the interferometer, which results from the insertion loss of phase modulators. This asymmetric loss leads to an unbalance of the amplitudes of the two generated pulses as illustrated in Fig.\;\ref{setup}. We describe here the general outline of the protocol structure. Since we are dealing with the asymptotic key rate in this article, we omit any detail that would be relevant only for a finite-size analysis of the protocol.

\begin{enumerate}
    \item \textit{State preparation}: Alice prepares a phase-randomised coherent state with mean photon number $|\alpha|^2$ where $\alpha\in\mathbb{C}$ and chooses a random phase $\phi_x$ from the set $\{0,\frac{\pi}{2},\pi,\frac{3\pi}{2}\}$ with equal probabilities in each round. Alice also sends a small portion of decoy coherent states with different mean photon numbers $\{|\alpha_i|^2:\forall \alpha_i\in\mathbb{C}\}_{i\in\mathbb{N}}$.
    
    \item \textit{Measurement}: Once Bob receives the signal state, he chooses a random phase $\phi_B$ from the set $\{0,\frac{\pi}{2}\}$ with equal probabilities and records all events coming from the two detectors at any of the three time slots. A click is termed ``outside" if it is not in the 2nd (middle) time slot.
    
    \item \textit{Testing}: After repeating steps 1 \& 2 for many times, Alice and Bob jointly announce a random subset of their data (including events coming from decoy states) and decide whether they should abort or proceed with the rest of the protocol.
    
    \item \textit{Announcement, sifting and post-selection}: For each round, Alice announces the basis to be ``even" if she picks her phase from $\{0,\pi\}$ or she announces ``odd" if her phase is in $\{\frac{\pi}{2},\frac{3\pi}{2}\}$. Bob announces ``even" if he picks $\phi_B = 0$ or ``odd" if $\phi_B = \frac{\pi}{2}$. In addition to basis announcements, Bob also announces ``discard" for events that have only outside clicks or no click. Alice keeps the $\phi_x$'s only for the rounds where Bob did not announce "discard" and where her bases match with Bob's. Bob keeps a detection event if his basis matches Alice's and the event is not to be discarded.
    
    \item \textit{Direct reconciliation key map}: Alice maps $\phi^{(j)}_x$ in the $j$-th kept rounds to the $j$-th bit $z_j$ of the raw key as
    \begin{equation}
    z_j = \begin{cases}
    0, \text{\hspace{5pt}if }\phi^{(j)}_x = 0, \pi/2,\\
    1, \text{\hspace{5pt}if }\phi^{(j)}_x = \pi, 3\pi/2.
    \end{cases}
    \end{equation}
    
    \item \textit{Error correction and privacy amplification}: Alice and Bob perform standard error correction so that Bob also obtains a copy of the key map register. They then proceed with a privacy amplification protocol to obtain a shared secret key.
\end{enumerate}
	
We point out that our method generalises to any asymmetric basis choice (i.e. probabilities of choosing ``even" and ``odd" bases are not equal). It was shown in Ref. \cite{AsymmetricBasisChoice} that the probability of choosing one basis can be set arbitrarily close to 1 without affecting the asymptotic security analysis. Note that the formalism described here would also allow one to consider the reverse reconciliation approach, where in step 5 of the protocol Bob performs a key map instead of Alice. Then, Alice and Bob would have to swap their respective roles in step 6.

\section{Mathematical Model of the protocol}\label{MathModel}
\subsection{Optical Models}
\begin{figure}[!htb]
    \centering
    \includegraphics[width=0.85\linewidth]{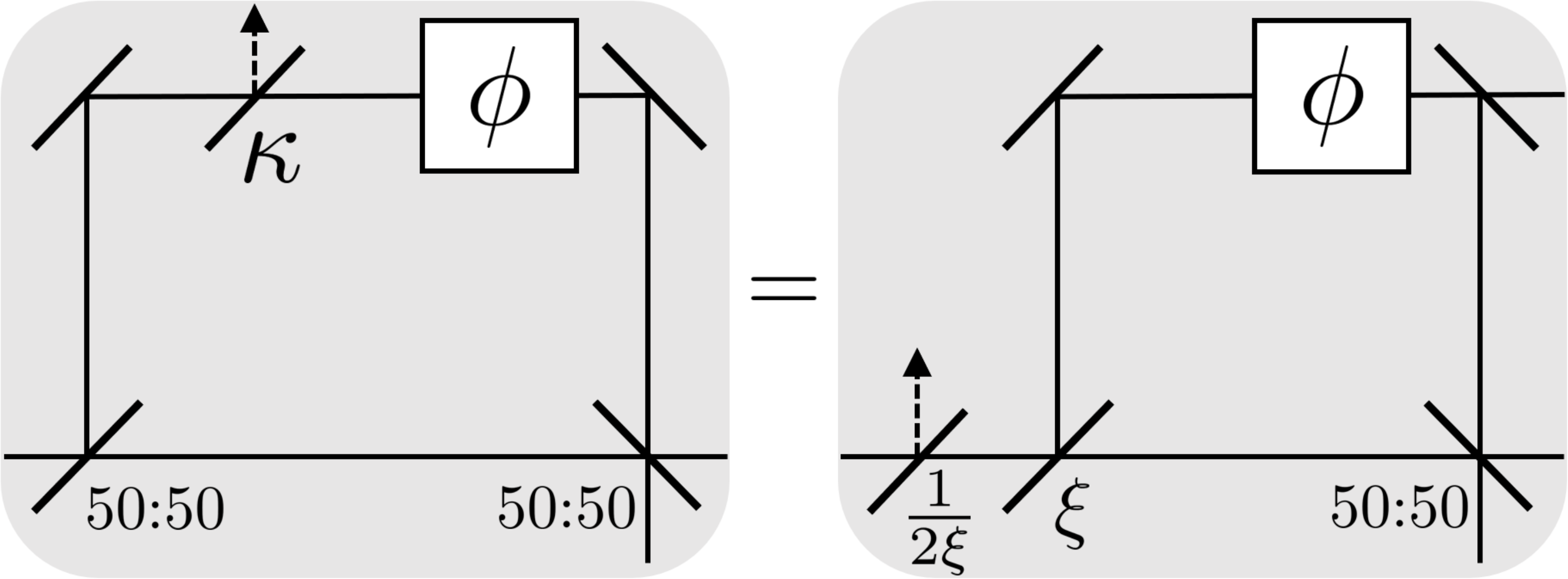}
    \caption{Equivalence relationship between a lossy phase modulator in the encoding device and an uneven BS with transmissivity $\frac{1}{2\xi}$ followed by another uneven BS with transmissivity $\xi$ and a perfect phase modulator, where $\xi = \frac{1}{1+\kappa}$ \cite{Agnes}.}
    \label{hardwareFix}
\end{figure}

We start by identifying two equivalent optical models for the Mach-Zehnder component that appears in both Alice's and Bob's apparatus. The descriptions for the two models are illustrated in Fig.\;\ref{hardwareFix}. Instead of having the loss in one arm of the interferometer, the equivalent model places a loss element in front of the Mach-Zehnder component, which then has an asymmetric beam splitter at the entry \cite{Agnes}. 

This replacement picture tells us that Alice's loss can be absorbed into the rescaled amplitude of the incoming single laser pulse, whereas Bob's loss can be absorbed into the channel's action. 

\subsection{State preparation} \label{StatePrep}
We use the source-replacement scheme \cite{SourceReplacementScheme,EntanglementPrecondition} to represent a prepare-and-measure scheme with an entanglement-based scheme. Since Alice's signal state is mixed, we will introduce a purifying ``shield" system that will be left behind in the source so that the existing source-replacement framework can be applied. We will provide a detailed description of the entangled pure state prepared by Alice below.

To prepare the output signal state, Alice's laser first creates a phase-randomised coherent state
\begin{equation}
    \sigma_\text{in}(2\alpha) = \int_0^{2\pi} \frac{d\theta}{2\pi}|2\alpha e^{i\theta}\rangle\langle 2\alpha e^{i\theta}| = \sum_{n=0}^\infty p_n(2\alpha)|n\rangle\langle n|,
\end{equation}
where $p_n(\beta) = e^{-|\beta|^2}\frac{|\beta|^{2n}}{n!}$ is the Poissonian distribution in photon number $n$. She then sends it through her encoding device set at a phase $\phi_x$ which outputs a time-bin signal with two modes,
\begin{equation}
    \sigma_x(\alpha) = \int_0^{2\pi} \frac{d\theta}{2\pi}|\psi^\theta_x(\alpha)\rangle\langle\psi^\theta_x(\alpha)|,
\end{equation}
where $|\psi^\theta_x(\alpha)\rangle = |\alpha e^{i\theta}, \sqrt{\kappa}\;\alpha e^{i(\theta-\phi_x)}\rangle$.

In the following steps, we will express the state $\sigma_x(\alpha)$ in a two-mode Fock basis $\{|s^x_n(\xi)\rangle\}$ which is defined later in Eqn. (\ref{snx}). Let $\widetilde{a}_1^{\;\dagger}$ and $\widetilde{a}_2^{\;\dagger}$ be the creation operators of the two output time modes of the signal. We define a rescaled amplitude $\widetilde{\alpha}\coloneqq \alpha\sqrt{1+\kappa} = \alpha/\sqrt{\xi}$ with the definition $\xi \coloneqq \frac{1}{1+\kappa}$
and a new mode creation operator
\begin{flalign}
    \widetilde{a}^{\;\dagger}_{\theta,x} &\coloneqq \frac{1}{\widetilde{\alpha}} (\alpha e^{i\theta}\;\widetilde{a}_1^{\;\dagger}+ \sqrt{\kappa}\;\alpha e^{i(\theta-\phi_x)}\;\widetilde{a}_2^{\;\dagger})\\
    &= \frac{e^{i\theta}}{\sqrt{1+\kappa}}(\widetilde{a}_1^{\;\dagger}+ \sqrt{\kappa}\;\alpha e^{-i\phi_x}\;\widetilde{a}_2^{\;\dagger})\\
    &= e^{i\theta}(\sqrt{\xi}\;\widetilde{a}_1^{\;\dagger}+\sqrt{1-\xi}\;e^{-i\phi_x}\;\widetilde{a}_2^{\;\dagger}).
\end{flalign}
We define a set of two-mode Fock states for $n\in\mathbb{N}$ as
\begin{flalign}
    |s^x_n(\xi)\rangle &= \frac{1}{\sqrt{n!}}(\widetilde{a}_{\theta=0,x}^{\;\dagger})^n|0\rangle \\
    &= \sum_{k=0}^n\sqrt{{n}\choose{k}}\xi^{\frac{n-k}{2}}(1-\xi)^\frac{k}{2}e^{-ik\phi_x}|n-k,k\rangle, \label{snx}
\end{flalign}
The state $|\psi^\theta_x(\alpha)\rangle$ can be rewritten in the new basis as
\begin{equation}
    |\psi^\theta_x(\alpha)\rangle = e^{-\frac{|\widetilde{\alpha}|^2}{2}} \sum_{n=0}^\infty \frac{\widetilde{\alpha}^{\;n}}{n!}(\widetilde{a}_{\theta,x}^{\;\dagger})^n|0\rangle = e^{-\frac{|\widetilde{\alpha}|^2}{2}} \sum_{n=0}^\infty \frac{(\widetilde{\alpha}e^{i\theta})^{n}}{\sqrt{n!}}|s^x_n(\xi)\rangle
\end{equation}
which is a coherent state with amplitude $\widetilde{\alpha}$. The phase-randomised signal state is therefore a Poissonian mixture of the new Fock states as in
\begin{equation}
    \sigma_x(\alpha) = \sum_{n=0}^\infty p_n(\widetilde{\alpha})|s^x_n(\xi)\rangle\langle s^x_n(\xi)|.
\end{equation}

Since the signal state $\sigma_x(\alpha)$ is mixed, Alice can purify the state by introducing an ancillary system $A_S$ such that the following is a pure state
\begin{equation}
    |\sigma_x(\alpha)\rangle_{A_SA'} = \sum_{n=0}^\infty \sqrt{p_n(\widetilde{\alpha})}\;|n\rangle_{A_S}\otimes|s^x_n(\xi)\rangle_{A'} \;\;,
    \label{sigmax}
\end{equation}
where the register $A'$ is the signal system. Note that the probability $p_n(\widetilde{\alpha})$ is independent of Alice's choice $x$.

We can thus summarise the source description as Alice preparing an entangled pure state
\begin{equation}
    |\Psi\rangle_{AA_SA'} = \sum_x \sqrt{p_x}\;|x\rangle_A\otimes|\sigma_x(\alpha)\rangle_{A_SA'} \;\;, \label{AliceFullState}
\end{equation}
where $\{|x\rangle_{A}\}_{x=0,...,3}$ is an orthonormal basis of Alice's register $A$ for $x$ corresponding to the phase $\phi_x = \frac{\pi}{2}x$ and $p_x =\frac{1}{4}$ for all $x\in\{0,1,2,3\}$. Note that registers $A$ and $A_S$ are private to Alice, and Eve only has access to the signal system $A'$. We call the purifying system $A_S$ a ``shield" system for it to be inaccessible to Eve (i.e. Eve only gets the mixed state $\sigma_x(\alpha)$ but not the pure state $|\sigma_x(\alpha)\rangle_{A_SA'}$).

\subsection{Measurements}\label{POVMSec}
In the prepare-and-measure scheme, the action of Alice randomly choosing the phase $\phi_x$ in the signal state is equivalent to a measurement on $|\Psi\rangle_{AA_SA'}$ with POVM $\{|x\rangle\langle x|_A\}_{x=0,...,3}$. Alice's measurement can be performed before or after Bob performs his measurement.

We start out by describing the POVM of Bob's measurement assuming ideal devices, especially without dark counts of the detectors. We will later on derive the POVM of devices with specified dark counts. To characterise all of Bob's possible measurement outcomes, we construct his POVM using the creation and annihilation operators for six optical modes arriving at 3 different time slots and at 2 detectors. Ignoring global phases, the six annihilation operators of a fixed phase $\phi_B$, which correspond to the six ``click" locations depicted in Fig. \ref{setup}, are
\begin{flalign}
    b_1 &= b_4 \rightarrow \sqrt{\frac{\xi}{2}}\;a_1 \;, \\ 
    b_{3} &= b_6 \rightarrow \sqrt{\frac{1-\xi}{2}}\;a_2 \;, \\
    b_{2,\phi_B} &\rightarrow \sqrt{\frac{1-\xi}{2}}\;a_1 - e^{i\phi_B}\sqrt{\frac{\xi}{2}}\;a_2 \;, \label{b2phiB}\\ b_{5,\phi_B} &\rightarrow \sqrt{\frac{1-\xi}{2}}\;a_1 + e^{i\phi_B}\sqrt{\frac{\xi}{2}}\;a_2 \;, \label{b5phiB}
\end{flalign}
where $a_1$ and $a_2$ are annihilation operators of the two incoming time modes of the signal. 

Since $b_1 = b_4$ and $b_{3} = b_6$, the POVM elements corresponding to click events at 1 and 4 (3 and 6) are the same. Hence, each pair can be combined into a single time-mode annihilation operator. The corresponding operators for the two pairs are
\begin{equation}
    b_{t_1} \rightarrow \sqrt{\xi}\;a_1 \;, \text{\hspace{10pt}}b_{t_3} \rightarrow \sqrt{1-\xi}\;a_2 \label{bt_1t_3}
\end{equation}
where $t_1$ and $t_3$ denote the 1st and 3rd time slots in Fig \ref{setup}. This is equivalent to coarse-graining the outside-only click POVM elements and outcome probabilities but without losing information about the relative phase, $\phi_x-\phi_B$. This reduces the redundancy in constraints for the optimisation which will be described in Sec. \ref{OPTSection}.

As Bob's measurement outcomes consist of all combinations of click events at different time slots, detectors, and basis choices, his POVM elements are obtained by summing weighted projectors of all possible states that could lead to a particular click pattern. Based on the fact that Bob uses threshold detectors for detection, all POVM elements are block-diagonal in total photon number basis \cite{SquashingPaper1,SquashingPaper2}.

These allow the construction of Bob's POVM elements in terms of the modes impinging on the detectors by first restricting to the $n$-total photon subspace of Bob's entire system, and defining the following operators corresponding to different click events: 
\begin{itemize}
    \item no-click: (for $n=0$) 
    \begin{equation}
        F_{0}^{\phi_B} = p(\phi_B)|0\rangle\langle0|,
        \label{vacuum}
    \end{equation} 
    \item single-click: (for $n\geq1$) 
    \begin{equation}
        F^{n,\phi_B}_{i_1} = p(\phi_B)\frac{1}{n!}(b_{i_1}^\dagger)^n|0\rangle\langle0|b_{i_1}^n \;, \label{Eq:singleclick}
    \end{equation} 
    \item double-click: (for $n\geq2$) 
    \begin{equation}
        F^{n,\phi_B}_{i_1,i_2} =p(\phi_B)\sum_{k=1}^{n-1}\frac{(b_{i_1}^\dagger)^{n-k}(b_{i_2}^\dagger)^k|0\rangle\langle0|b_{i_1}^{n-k}\;b_{i_2}^k}{(n-k)!\;k!} \;,  \label{Eq:doubleclick}
    \end{equation}
    \item triple-click: (for $n\geq3$)
    \begin{equation}
        F^{n,\phi_B}_{i_1,i_2,i_3} =p(\phi_B)\sum_{k=1}^{n-2}\;\sum_{j=1}^{n-k-1}|\beta_3(n,j,k)\rangle\langle \beta_3(n,j,k)|
    \end{equation}
    with $|\beta_3(n,j,k)\rangle = \frac{(b_{i_1}^\dagger)^{n-k-j}(b_{i_2}^\dagger)^k(b_{i_3}^\dagger)^j|0\rangle}{\sqrt{(n-k-j)!\;k!\;j!}}$ ,
    \item all-click: (for $n\geq4$) 
    \begin{equation}
        F^{n,\phi_B}_\text{ac} =p(\phi_B)\sum_{k=1}^{n-3}\;\sum_{j=1}^{n-k-2}\;\sum_{l=1}^{n-k-j-1}|\beta_4(n,j,k,l)\rangle\langle \beta_4(n,j,k,l)|
        \label{allclick}
    \end{equation}
    with $|\beta_4(n,j,k,l)\rangle = \frac{(b_{i_1}^\dagger)^{n-k-j-l}(b_{i_2}^\dagger)^k(b_{i_3}^\dagger)^j(b_{i_4}^\dagger)^l|0\rangle}{\sqrt{(n-k-j-l)!\;k!\;j!\;l!}}$ ,
\end{itemize}
where $p(\phi_B)$ is the probability of choosing the phase $\phi_B$ and $b_{i_\mu}^\dagger\in\{b_{t_1}^\dagger, b_{2,\phi_B}^\dagger, b_{5,\phi_B}^\dagger, b_{t_3}^\dagger\}$ are the mode creation operators for a fixed phase $\phi_B$, with $b_{i_\mu}^\dagger\neq b_{i_\nu}^\dagger$ for all $\mu \neq \nu$ and $\mu,\nu\in\{1,2,3,4\}$. We can express Bob's POVM elements in terms of the incoming modes, $a_1$ and $a_2$, by substituting the final modes with Eqns. (\ref{b2phiB}) -- (\ref{bt_1t_3}).

To obtain Bob's POVM elements for the full Hilbert space, one simply sums over all contributions from all photon number subspaces to get
\begin{equation}
    F_k = \sum_{n=0}^\infty F_k^n \;,
\end{equation}
where $k$ labels the 16 possible click patterns (Bob's measurement outcomes) in each of the two measurement bases. For $n$ to be less than the minimum photon number to trigger the click event $k$, $F_k^n$ is a zero operator. If $k$ is the no-click event, $F_k^n$ is a zero operator for all $n\geq1$.

To reduce the number of linearly dependent POVM elements for better numerical performance in calculating key rates \footnote{As we will point out in Sec. \ref{FlagStateSection}, the number of POVM elements is related to the dimension of the flag-state subspace. If the two POVM elements are linearly dependent, they are essentially the same constraint for the convex optimisation problem in (\ref{OPT}) up to a scaling factor. Therefore, omitting either of the two elements will not affect the optimisation result, but the flag-state subspace dimension will reduce by one. As for all numerical optimisations, the smaller the dimension of the problem, the shorter the runtime.}, we combine the pairs of $\phi_B$-independent POVM elements of the two measurement bases into one by summing the two elements together. This reduces the cardinality of Bob's POVM from 32 to 28 since the following four click patterns: no-click, $t_1$-only, $t_3$-only, and $t_1\&t_3$ are basis-independent.

In a trusted dark-count scenario where dark counts are not controlled by Eve, we incorporate the effect of dark counts into Bob's POVM by applying a classical post-processing map, $\mathcal{P}$, on Bob's POVM elements $\{F_k\}$. The output of the map is a new POVM $\{P_k\}$ with each element corresponding to a linear combination of the original POVM such that $P_k = \sum_i \mathcal{P}_{k,i} \;F_i$ where $\mathcal{P}_{k,i}$ are the matrix elements of the linear map $\mathcal{P}$. We illustrate the action of the map $\mathcal{P}$ with the new POVM elements in Eqns. (\ref{Eq:CPP0})-(\ref{Eq:CPP25}). Since the map $\mathcal{P}$ acts the same on all photon-number subspaces, it also holds that 
\begin{equation}
    P_k^n = \sum_i \mathcal{P}_{k,i}\; F_i^n \;. \label{Eq:CPP_n}
\end{equation}
The map $\mathcal{P}$ models the effect of dark counts as a classical noise in the sense that for each detector and at each detection time window, a no-click event flips to a click event with probability, $p_d$. We can recover Bob's dark-count free POVM $\{F_k\}$ by setting the dark-count probability $p_d=0$ in the case with untrusted dark counts.

Overall, we obtain the joint POVM of Alice's and Bob's measurements $\{|x\rangle\langle x|_A \otimes P_k\}$ where $x\in\{0,...,3\}$ and $k\in\{1,...,28\}$ since Bob has 28 coarse-grained outcomes in total if no-click is included.

\section{Security proof techniques}\label{SecurityProofTech}

\subsection{Flag-state squashing model}\label{FlagStateSection}
In order to numerically compute the secure key rate, we need to reduce the dimension of Bob's state from infinite to finite so that numerical optimisation solvers can be used. Since Bob uses threshold detectors, his POVM elements are block-diagonal, so the qubit squashing model \cite{SquashingPaper1,SquashingPaper2,Agnes,VarunThesis} can be applied. However, by reassigning the multi-click events to single-click events randomly, the squashing model introduces additional qubit errors to the original data. Instead, the flag-state squashing model \cite{Yanbao} is used here to circumvent this problem.

We set a finite photon-number cutoff $N_B$ and define the $(n$\;$\leq$\;$N_B)$- and $(n$\;$>$\;$N_B)$-photon subspaces to be two Hilbert spaces containing Fock states of at most $N_B$ and at least $N_B+1$ photons respectively. The flag-state squashing map $\Lambda$ first projects Bob's state $\rho$ onto the two subspaces. It then applies an identity map to the projected state $\rho_{n\leq N_B}$ and measures the projected state $\rho_{n>N_B}$ with the POVM $\{P_k\}$ to give the squashed state
\begin{equation}
    \Lambda(\rho) = \begin{pmatrix}
    \rho_{n\leq N_B} & 0\\  0 & \sum_k \text{Tr}(P_k \;\rho_{n>N_B})|k\rangle\langle k|\end{pmatrix}.
    \label{flagstateSquash}
\end{equation}
Bob's corresponding flag-state squashed POVM elements are 
\begin{equation}
    \widetilde{P}_k = \left(\sum_{n=0}^{N_B} P_k^n\right) \oplus |k\rangle\langle k|,
    \label{flagPOVM}
\end{equation}
where $N_B$ is a finite-number photon cutoff and $k$ labels Bob's detection events. The joint POVM of Alice's and Bob's measurements in the flag-state squashing model is $\{|x\rangle\langle x|_A \otimes \widetilde{P}_k\}$ where $x\in\{0,...,3\}$ and $k\in\{1,...,28\}$.

Since the measurement channel acting on the $(n>N_B)$-photon subspace is entanglement breaking \cite{EntBreaking}, one needs to lower bound $\Tr(\Pi_{n\leq N_B}\;\rho)$ with Bob's measurement statistics to ensure that some entanglement between Alice and Bob is preserved in order for them to establish a secret key \cite{EntanglementPrecondition}. For trusted dark counts, we show in Appendix \ref{App:LowerBound} that the lower bound for the weight of the $(n\leq N_B)$-photon signal subspace conditioned on Alice choosing signal $x$ is given by
\begin{gather}
    p(n\leq N_B|x) \geq 1 - \frac{p(\text{cc}|x) - p(\text{cc}|0)}{p_{\text{min}}(\text{cc}|N_B+1) - p(\text{cc}|0)} \;, \label{lowerbound}\\
    p(\text{cc}|0) = 1 - (1-p_d)^2 [1 + p_d (1-p_d)^2 (2-p_d)], \label{pcc|0}\\
    p_{\text{min}}(\text{cc}|n) = 1 - (1-p_d)^2 \xi^{n} - (1-p_d)^4 (1-\xi)^{n}, \label{p_min_cc|n}
\end{gather}
where the conditional cross-click probability, $p(\text{cc}|x)$, is the sum of the observed probabilities of all events excluding no-click events, events with clicks only in time slot $t_2$ (inside-only), and events with clicks only in time slots $t_1, t_3$ (outside-only) given that Alice picks signal $x$.  We also show in Appendix \ref{App:LowerBound} that the bound in (\ref{lowerbound}) is always tighter than the dark-count free bound ($p_d=0$) derived by Narashimhachar \cite{VarunThesis}, so we could also obtain a lower bound of the secure key rate using that dark-count free bound. For untrusted dark counts, one simply has to use that bound.

\subsection{Decoy state \& decomposition of key rate formula} \label{Sec:decoy}
In this article, we prove the security of the protocol against any collective attack. Since the signal states and measurements are permutation invariant between different rounds, the quantum de Finetti theorem \cite{iidRenner2007} or the postselection technique \cite{PostselectTech} can be applied to uplift our security statement to the security against coherent attacks, which will both lead to the same asymptotic key rate. From that we obtain a composable $\epsilon$-security proof \cite{rennerThesis} of the protocol under Eve's general attacks with the same asymptotic key rate as under the collective attack.

Let $R$ be the key register held by Alice in direct reconciliation, $E$ be Eve's quantum and classical register, $B$ be Bob's quantum register, $\overline{B}$ and $\widetilde{B}$ be Bob's classical registers for his measurement outcomes and announcements respectively. The Devetak-Winter formula \cite{DevetakWinter} for asymptotic secure key rate can be expressed as
\begin{equation}
    R_\infty = p_\text{pass}[\min_{\rho\in\mathbf{S}} H(R|E) - H(R|\overline{B})], \label{DWFormula}
\end{equation}
where $p_{\text{pass}}$ is the probability of passing the sifting and post-selection steps, $\mathbf{S}$ is the set of all density matrices that satisfy Alice's and Bob's joint statistics.

The key rate formula (\ref{DWFormula}) can be converted into an alternative form, as shown in Refs. \cite{Coles2016,AdamNumerics}, using the relative entropy
\begin{equation}
    R_\infty = \min_{\rho_{AA_SB}\in\mathbf{S}}D(\mathcal{G}(\rho_{AA_SB})||\mathcal{Z}(\mathcal{G}(\rho_{AA_SB}))) - p_{\text{pass}}\;\delta_\text{EC} \;, \label{KeyRateFormula2}
\end{equation}
where $\mathcal{G}$ and $\mathcal{Z}$ are two maps that will be discussed below. The formula includes a privacy amplification (PA) term as the first term and an error correction term $\delta_\text{EC} = f_\text{EC}\;H(R|\overline{B})$ with a heuristic classical error-correction efficiency factor $f_\text{EC}\geq 1$.

The $\mathcal{G}$ map is a completely positive trace non-increasing map capturing the effects of measurements, sifting, post-selection and announcement on Alice's and Bob's joint state, which takes the form \cite{AdamNumerics,JiePRX}
\begin{equation}
    \mathcal{G}(\rho) = \sum_i K_i\; \rho\; K_i^\dagger \label{Gmap}
\end{equation}
with the Kraus operators of this protocol defined as
\begin{equation}
     K_0 = (|0\rangle_R\otimes|0\rangle\langle0|_A + |1\rangle_R\otimes|2\rangle\langle2|_A)\otimes\mathcal{F}_0^B\otimes |0\rangle_{\widetilde{B}} \;\;, \label{Kraus0}
\end{equation}
\begin{equation}
    K_1 = (|0\rangle_R\otimes|1\rangle\langle1|_A + |1\rangle_R\otimes|3\rangle\langle3|_A)\otimes\mathcal{F}_1^B\otimes |1\rangle_{\widetilde{B}} \;\;, \label{Kraus1}
\end{equation}
where  $\{|0\rangle_{\widetilde{B}}, |1\rangle_{\widetilde{B}}\}$ is Bob's basis announcement bit, $\mathcal{F}_j^B = \sqrt{\sum_{b\in\mathbf{K}} F_{b,\phi_B=\frac{\pi}{2}j}}$ and $\mathbf{K}$ denotes Bob's post-selected outcomes. The $\mathcal{Z}$ map captures the effect of the key map, and is given by
\begin{equation}
    \mathcal{Z}(\sigma_{RC}) = \sum_{j=0}^1 (|j\rangle\langle j|_R\otimes\mathbb{1}_C)\;\sigma_{RC}\; (|j\rangle\langle j|_R\otimes\mathbb{1}_C) \label{Zmap}
\end{equation}
with register $C$ encapsulates all registers except $R$.

Since Alice is sending a Poissonian mixture of Fock states, Eve can, in principle, perform a QND measurement on Alice's signal to learn its photon number without disturbing the signal itself. We show in Appendix \ref{PADecompApp} that as a direct consequence of this the state $\rho_{AA_SB}$ is block-diagonal in Alice's output photon number $\widetilde{n}$. Therefore, without loss of generality, we can restrict the minimisation in Eqn. (\ref{KeyRateFormula2}) to be taken over a smaller set $\mathbf{S}'=\{\rho_{AA_SB}\in\mathbf{S}: \rho_{AA_SB} = \sum_{\widetilde{n}=0}^\infty p_{\widetilde{n}}\;|\widetilde{n}\rangle\langle \widetilde{n}|_{A_S}\otimes \rho_{AB}^{\widetilde{n}}\}$ where $\{\rho_{AB}^{\widetilde{n}}\}$ are the normalised states conditioned on Alice sending out $\widetilde{n}$ photons. This allows one to split the PA term into a probabilistic combination of PA terms associated with different $\widetilde{n}$ as in
\begin{equation}
    R_\infty = \min_{\rho_{AA_SB}\in\mathbf{S}'}\sum_{\widetilde{n}=0}^\infty p_{\widetilde{n}}\; D(\mathcal{G}(\rho_{AB}^{\widetilde{n}})||\mathcal{Z}(\mathcal{G}(\rho_{AB}^{\widetilde{n}}))) - p_{\text{pass}}\; \delta_\text{EC}\;. \label{KeyRateS'}
\end{equation}
See Appendix \ref{PADecompApp} for the proof of the decomposition.

For our analysis, we assume a decoy-state scenario \cite{Decoy0,Decoy1,Decoy2}, which means that in addition to the usual signal states, Alice prepares also decoy states that are represented by dephased laser pulses with different intensity levels $|\alpha_i|^2$. More precisely, we assume for simplicity the infinite-decoy scenario, where a countably infinite number of decoy intensities are used so that a decoy data analysis can reveal to Alice and Bob the conditional probabilities of any observable, where the condition is with respect to Alice's output photon number $\widetilde{n}$.

These conditional probabilities constrain the feasible set of normalised states $\mathbf{S}_{\widetilde{n}}$ for each of Alice's output photon number $\widetilde{n}$ independently, which further restricts the minimisation in Eqn. (\ref{KeyRateS'}) to be taken over a smaller set $\mathbf{S}''=\{\rho_{AA_SB}\in\mathbf{S}:\rho_{AA_SB} = \sum_{\widetilde{n}=0}^\infty p_{\widetilde{n}}\;|\widetilde{n}\rangle\langle \widetilde{n}|_{A_S}\otimes \rho_{AB}^{\widetilde{n}},$  $\rho_{AB}^{\widetilde{n}}\in\mathbf{S}_{\widetilde{n}}$ $\forall$ ${\widetilde{n}}\in\mathbb{N}\} \subset\mathbf{S}'$. Given that the probability distribution $\{p_{\widetilde{n}}\}_{{\widetilde{n}}\in\mathbb{N}}$ is fixed by the intensity of the signal, the minimisation over $\mathbf{S}''$ can be pulled into the summation and split into minimisations over individual $\mathbf{S}_{\widetilde{n}}$, resulting in the following key rate formula
\begin{equation}
    R_\infty = \sum_{\widetilde{n}=0}^\infty p_{\widetilde{n}} \min_{\rho_{AB}^{\widetilde{n}}\in\mathbf{S}_{\widetilde{n}}}D(\mathcal{G}(\rho_{AB}^{\widetilde{n}})||\mathcal{Z}(\mathcal{G}(\rho_{AB}^{\widetilde{n}}))) - p_{\text{pass}}\; \delta_\text{EC}\;.
    \label{minSn}
\end{equation}
We remark that the inclusion of a finite number of decoy states would be a natural extension of this work, in which case the description of each set $\mathbf{S}_{\widetilde{n}}$ would depend on other sets $\{\mathbf{S}_{n'}: n'\neq \widetilde{n}\}$. Hence, a more careful treatment of the PA term would be needed.

The major benefit of breaking down the PA term into individual minimisations is to avoid the need of keeping the infinite-dimensional shield system $A_S$ in the argument of the optimisation as seen in Eqn. (\ref{KeyRateFormula2}). Instead of optimising over the set of infinite-dimensional states, we convert our problem into an infinite number of optimisations with finite-dimensional arguments. 

Notice that when Alice sends out vacuum (0 photons), Eve learns nothing about Alice's choice $x$, so each key bit $z\in\{0,1\}$ is equally likely to Eve, which implies that $H(R|E) = H(R) = 1$. Therefore, the first term in the summation in Eqn. (\ref{minSn}) is equal to $p_{\text{pass}}^{\widetilde{n}=0}$ which is the contribution from Alice sending out vacuum to the probability of passing sifting and post-selection.

By Klein's inequality, quantum relative entropy is non-negative, i.e. $D(A||B)\geq0$, for all positive semidefinite matrices $A,B\geq0$ such that $\Tr(A)\geq\Tr(B)$ \cite{watrous_2018}, so $D(\mathcal{G}(\rho_{AB}^{\widetilde{n}})||\mathcal{Z}(\mathcal{G}(\rho_{AB}^{\widetilde{n}})))\geq 0$ $\forall$ $\widetilde{n}\in\mathbb{N}$. Thus, omitting any terms in the summation will only reduce the total value on the right-hand side of Eqn. (\ref{minSn}). In fact, omitting an $\widetilde{n}$-photon term is the same as treating all $\widetilde{n}$-photon output signals as being tagged for which the encoded state is fully known to Eve. Since we can only optimise a finite number of terms in the infinite sum, we can truncate the infinite sum at $\widetilde{n}=N_A$ where $N_A$ is a positive finite integer to obtain a lower bound for the key rate. The choice of $N_A=1$ corresponds to the tagging as used in Refs. \cite{Agnes,Kiyoshi}. We then have the key rate expression as
\begin{equation}
    R_\infty \geq p_{\text{pass}}^{\widetilde{n}=0} + \sum_{\widetilde{n}=1}^{N_A} p_{\widetilde{n}} \min_{\rho_{AB}^{\widetilde{n}}\in\mathbf{S}_{\widetilde{n}}}D(\mathcal{G}(\rho_{AB}^{\widetilde{n}})||\mathcal{Z}(\mathcal{G}(\rho_{AB}^{\widetilde{n}}))) - p_{\text{pass}}\; \delta_\text{EC}\;.
    \label{KeyRateLowBound}
\end{equation}
This allows us to reduce the number of finite-dimensional optimisations from infinity to a finite number that corresponds to the limited computational resources available to us.

\subsection{The optimisation problem}\label{OPTSection}
The convex optimisation problem corresponding to each PA term in Eqn. (\ref{KeyRateLowBound}) can be formulated as
\begin{flalign}
    &\text{minimise } D(\mathcal{G}(\rho_{AB}^{\widetilde{n}})||\mathcal{Z}(\mathcal{G}(\rho_{AB}^{\widetilde{n}})))\nonumber\\
    &\text{subject to}\nonumber\\
    & \text{\hspace{15pt}} \text{Tr}[(|x\rangle\langle x|_A  \otimes \widetilde{P}_k)\;\rho_{AB}^{\widetilde{n}}] = p(x,k|\widetilde{n}),\nonumber\\
    & \text{\hspace{15pt}} \text{Tr}[(|x\rangle\langle x|_A\otimes\Pi_{n\leq N_B})\; \rho_{AB}^{\widetilde{n}}] \geq p(x)\;p^\text{min}_{n\leq N_B|x} \;,\nonumber\\
    & \text{\hspace{15pt}} \Tr_B(\rho_{AB}^{\widetilde{n}}) = \frac{1}{p_{\widetilde{n}}}\text{Tr}_{A_SA'}[(|\widetilde{n}\rangle\langle \widetilde{n}|_{A_S}\otimes\mathbb{1}_{A'})\;|\Psi\rangle\langle\Psi|_{AA_SA'}],\nonumber\\
    & \text{\hspace{15pt}} \text{Tr}(\rho_{AB}^{\widetilde{n}}) = 1,\nonumber\\
    & \text{\hspace{15pt}} \rho_{AB}^{\widetilde{n}} \geq 0.
    \label{OPT}
\end{flalign}

The first line in the constraints demands the shared state $\rho_{AB}^{\widetilde{n}}$ conditioned on Alice sending out $\widetilde{n}$ photons to satisfy Alice's and Bob's joint measurement outcome probabilities conditioned on $\widetilde{n}$, which are obtained from the infinite-decoy analysis. The second line lower bounds the weight of $\rho_{AB}^{\widetilde{n}}$ in the $(n\leq N_B)$-photon subspace by Eqn. (\ref{lowerbound}). The third line demands that Alice's reduced density matrix is unchanged. The last two lines ensure that $\rho_{AB}^{\widetilde{n}}$ is a valid, normalised density matrix.

\subsection{Implementation of numerical security analysis} \label{NumericalImplementation}

Following the procedure in Ref. \cite{AdamNumerics}, the suboptimal solutions to the convex optimisation problem (\ref{OPT}) for $1\leq \widetilde{n}\leq N_A$ are obtained numerically using the MATLAB optimisation package CVX and the Frank-Wolfe algorithm \cite{FrankWolfe}. These suboptimal solutions infer the upper bound for the individual privacy amplification terms in Eqn. (\ref{KeyRateLowBound}). A linearisation of each of the optimisation problems at its suboptimal solution results in a primal semidefinite programming (SDP) problem which can be further converted into a dual SDP problem. Using the CVX numerical solver again, the dual suboptimal solutions for $1\leq \widetilde{n}\leq N_A$ provide a reliable lower bound on the whole privacy amplification term.

Solving the convex optimisation problem is computationally demanding in terms of time and memory even if the flag-state squashing model is applied to reduce the dimension of the matrix variables $\rho_{AB}^{\widetilde{n}}$. One can further utilise the structure of the flag-state squashed state as described in Eqn. (\ref{flagstateSquash}) to reduce the number of complex variables in the allowed matrices $\rho_{AB}^{\widetilde{n}}$. Bob's flag-state squashed POVM elements also enable us to split multiplications between constraint matrices and the state variable $\rho^{\widetilde{n}}_{AB}$. In addition, the objective function in (\ref{OPT}) can be evaluated much faster if the computation is restricted only to the non-zero subspaces in the images of the maps $\mathcal{G}$ and $\mathcal{Z}$. With these three techniques, we managed to reduce the computation time of the convex optimisation by a significant amount. See Appendix \ref{App:SpeedupMain} for the technical details.

We utilise the fact that the optimisation problem specified in (\ref{OPT}) is independent of the mean photon number $|\alpha|^2$ of Alice's phase-randomised coherent state because the minimisations in Eqn. (\ref{minSn}) are over each set $\mathbf{S}_{\widetilde{n}}$ separately. In other words, the choice of $|\alpha|^2$ only affects the photon number distribution $\{p_{\widetilde{n}}\}$ and the error-correction term $\delta_\text{EC}$ in the key rate formula (\ref{KeyRateLowBound}). Therefore, we can maximise the key rate lower bound over the signal intensity $|\alpha|^2$ efficiently once we have the dual suboptimal solutions since the error-correction term can be directly calculated from the observables of the corresponding simulation.

\section{Simulation of experiments}\label{SimulateExperiment}
In the absence of experimental data, we have to perform a simulation of an experiment to obtain realistic probability distributions which replace the experimental data as input of our security analysis. Note that the details of the simulation model are independent of the actual security proof.

\subsection{Channel simulations \& detection efficiency}

We simulate the quantum channel between Alice and Bob with a loss-only channel which is essentially an uneven beam splitter. We also assume that both detectors of Bob have equal detection efficiency $\eta_{\text{det}}$, where each detector can be modelled as a beam splitter with a transmission rate $\eta_{\text{det}}$ followed by an ideal detector. In this simple model, a single parameter $\eta$ which we call the total transmissivity describes the combined loss caused by the following three effects: the inefficiency in the process of coupling the signal light to the optical fibre, the absorption and scattering processes of light in transmission through the fibre, and the detection efficiency of Bob's threshold detectors. 

We also investigate the case where we assume the detection efficiency $\eta_{\text{det}}$ to be outside of Eve's control, as a trusted, characterised loss element of the receiver. In that case, we keep the beam splitter with transmissivity $\eta_{\text{det}}$ in Bob's apparatus, which in turn modifies the POVM elements described in Sec. \ref{POVMSec}. Bob's POVM with known detection efficiency can be obtained with a similar approach used in Ref. \cite{Yanbao}.

\subsection{Dark counts} \label{Sec:DarkCounts}
To simulate our statistics when dark counts are present, we generate the outcome probabilities with Bob's classically post-processed POVM described in Sec. \ref{POVMSec} and Appendix \ref{App:LowerBound}, which is associated with a dark-count probability, $p_d$, for each detector and at each detection time window.

If dark counts are assumed to be trusted in the sense that they are not in Eve's control, we use the classically post-processed flag-state POVM $\{\widetilde{P}_k\}$ as the constraint matrices in the optimisation problem (\ref{OPT}) to calculate the privacy amplification term. This approach guarantees the optimisation problem to be feasible since measurement probabilities correspond directly to a quantum state in the simulation.

However, if we consider untrusted dark counts, that is, if we pessimistically attribute the effect of dark count noise to Eve, the flag-state POVM of dark-count free detectors is used as the optimisation constraint matrices instead. Note that unlike the existence of a physical model for pulling out the equal detection efficiency into the channel, this approach is not covered by any physical equivalence model that allows one to outsource the dark counts to Eve. Therefore, it is possible that no quantum states could have led to the classically post-processed statistics if the measurement is assumed to be dark-count free. In that case, the optimisation problem becomes infeasible due to unphysical constraints. This is what we encounter in some parameter regime of our calculation, as we will point out in the next section.

\section{Key rates}\label{ResultsSec}

Before diving into our main results, we start by stating the parameters used throughout this section. We set Bob's flag-state photon number cutoff to be $N_B=4$ so that the PA term can be computed within a reasonable amount of time. The maximum number of terms kept in the PA summation in Eqn. (\ref{KeyRateLowBound}) is set to be $N_A=3$ since we observe that the key rate in the low-loss regime does not improve even if we keep more than 3 terms. Furthermore, we set the dark count probability to be $p_d=8.5\times10^{-7}$ and the error-correction efficiency to be $f_\text{EC}=1.22$ as quoted in Ref. \cite{GYSexperiment}.

\begin{figure}[!htb]
\renewcommand{\arraystretch}{0}
\centering
\begin{tabular}{@{}c@{}}
    \includegraphics[width=\linewidth]{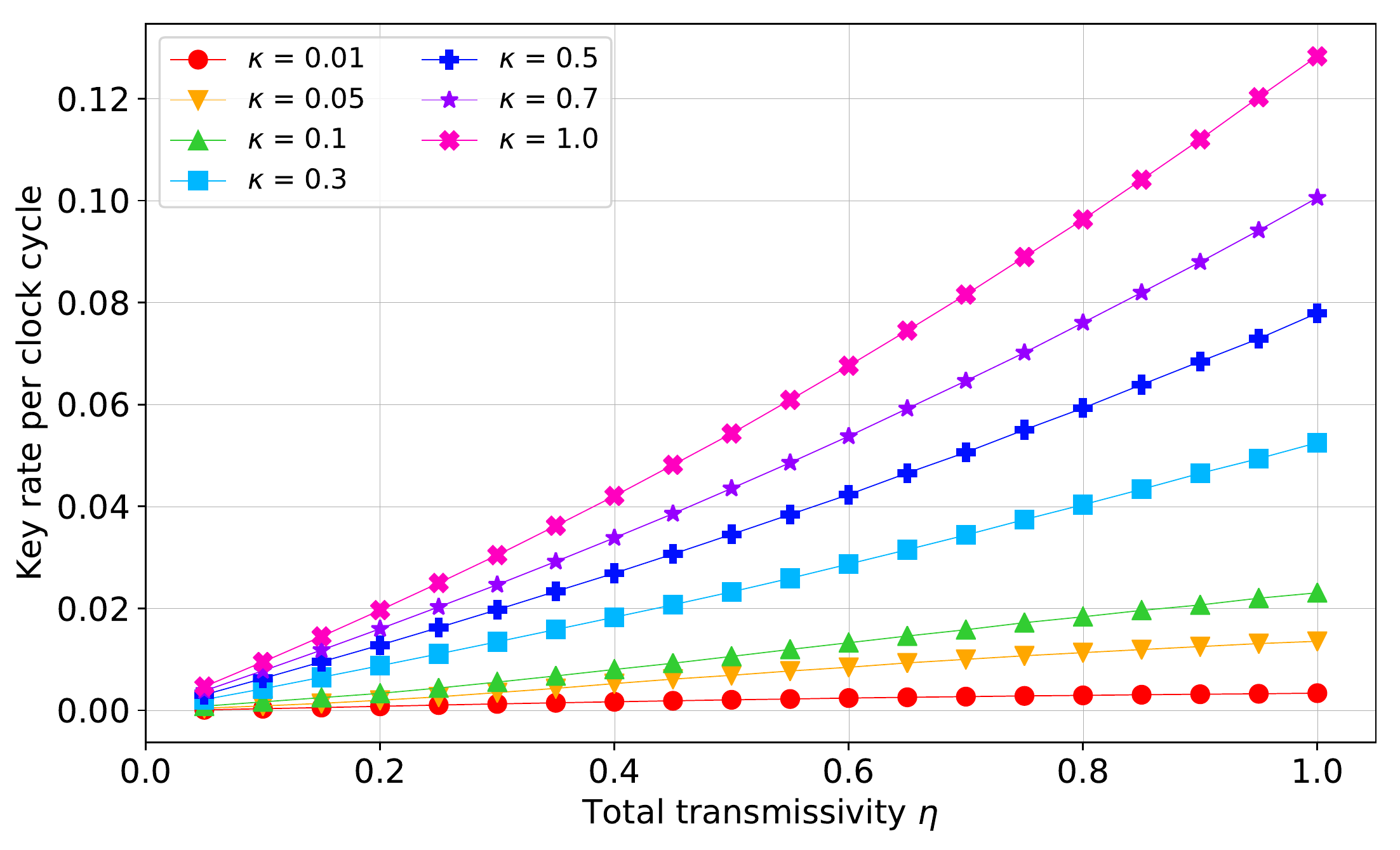}\\
    \text{(a)}\\
    \includegraphics[width=1.1\linewidth]{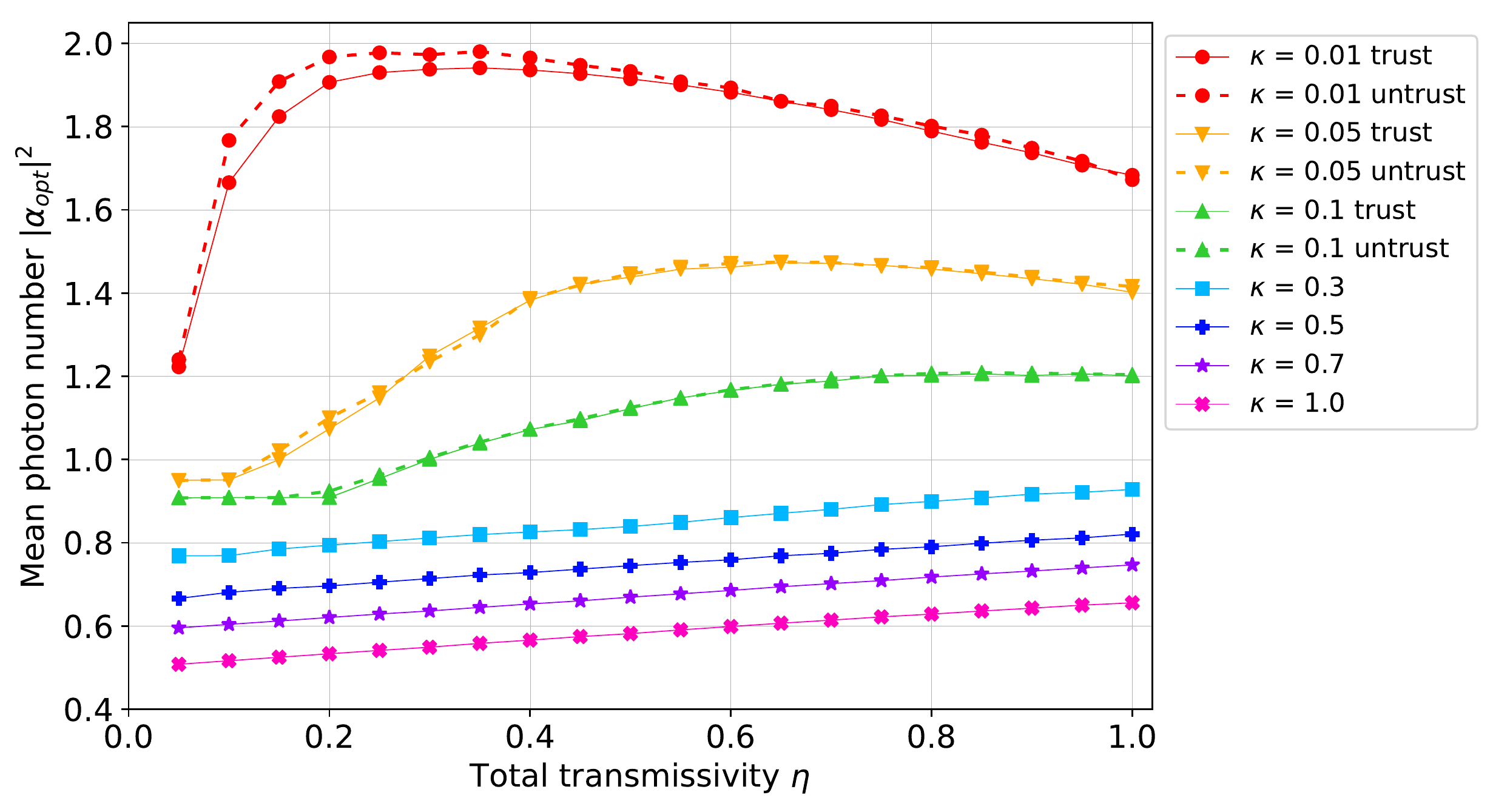} \\
    \text{(b)}
  \end{tabular}
\caption{(a) Our optimal lower bounds and (b) the corresponding mean photon numbers for secure key rates per clock cycle for both trusted (solid lines) and untrusted dark counts (dotted lines) versus total transmissivity $\eta$. For clarity, we omit labelling the lines for trusted and untrusted dark counts in the cases where the two lines are indistinguishable.}\label{OptimalKeyRate}
\end{figure}

In Fig.\;\ref{OptimalKeyRate}(a), we present lower bounds for the secure key rates per clock cycle corresponding to different values of the phase-modulator transmissivity $\kappa$ and the total transmissivity $\eta$ in the two scenarios with trusted and untrusted dark counts. The total transmissivity $\eta$ captures both the transmission efficiency of the loss-only channel and the detection efficiency of Bob's detectors. We obtain these bounds by maximising the lower bounds for key rates over the mean photon number $|\alpha|^2$ as specified in Sec. \ref{NumericalImplementation}. The optimal $|\alpha|^2$ for each point in Fig. \ref{OptimalKeyRate}(a) are shown in Fig. \ref{OptimalKeyRate}(b).

Let us expand on the infeasibility issue with untrusted dark counts mentioned in Sec.\;\ref{Sec:DarkCounts}. In the high-loss regime where the total transmissivity $\eta\leq0.2$, the optimisation problem for some parameters becomes infeasible meaning that no physical states can satisfy the constraints that are imposed by observed statistics. This is a somehow surprising observation since many previous security analyses (e.g. \cite{Agnes,Kiyoshi,Tagging1,ScaraniReviewPaper}) assume dark counts to be untrusted but did not encounter any issue with infeasible constraints. Most of these analyses use coarse-grained statistics (e.g. bit/phase error rate) to bound Eve's knowledge. However, the use of refined statistics in our optimisation constraints poses more stringent conditions on the feasible set which makes it less robust against infeasibility issues. Therefore, at least when infeasibility is detected, we cannot outsource the dark counts simulated by a classical noise model entirely to Eve as previous literature did. In the case of having infeasible data, we allow the numerical solver to relax the satisfiability of constraints in the sense that we are enlarging the search set to the degree where it is feasible. Due to large constraint violations and a minimisation over an enlarged search set, we expect the key rate lower bound obtained by this method to be much lower than the true value. As for the feasible cases, Fig.\;\ref{OptimalKeyRate}(a) shows that turning dark counts from untrusted to trusted increases the key rates. In the remaining of this section, if we make statements about the key rates without mentioning whether dark counts are trusted or untrusted, then the statement applies to both cases.

In the design view of a QKD security analysis, the goal is to optimise over all parameters and find the optimal setting of the experimental setup. Here, we seek the optimal asymmetric transmission parameter $\kappa$ and the corresponding optimal signal intensity $|\alpha|^2$ that gives the highest key rate at different total transmissivity $\eta$. We see that the smaller the value of $\kappa$, the lower the key rates in Fig.\;\ref{OptimalKeyRate}(a) because Alice would need to send more photons (as one can see from Fig. \ref{OptimalKeyRate}(b)) in order to maintain an adequate proportion of middle-click detection events, which allow Bob to infer the relative phase $\phi_x-\phi_B$. Therefore, one should always aim at reducing the loss at the phase modulator in order to increase the overall key rate.

To elaborate more on the optimality of the intensities in Fig.\;\ref{OptimalKeyRate}(b), we point out the two competing factors for using more photons in the signal. First, sending higher intensity signals causes more photons to pass through Eve's domain, which allows her to gain more information about the signal, thereby reducing the key rate. Second, as more information can be transmitted from Alice to Bob via multi-photon signals, the key rate may increase if the cost of error correction increases less than the information gain by Eve.

These two factors pull the key rate into opposite directions, so there is an optimal point for the key rate to be maximised, of which the corresponding optimal mean photon number is shown in Fig.\;\ref{OptimalKeyRate}(b). These values appear to be higher than the optimal values for the key rates in \cite{Kiyoshi}. This indicates that some multi-photon signals carry useful information from Alice to Bob of which Eve does not possess full knowledge, and hence favours signals with higher intensity.

At this point, we would like to compare our results with previous results in \cite{Agnes,Kiyoshi} which both contain valid security proofs that make use of the single-photon components only. Note that although the technical analysis of \cite{Agnes} is correct, the conclusion that the key rate of the unbalanced BB84 protocol will be overestimated if one blindly uses the security analysis of a balanced protocol is not. While \cite{Agnes} has shown that the key rate for unbalanced signals is lower than that for balanced ones, the authors of \cite{Kiyoshi} correctly point out that the drop in key rate is due to a smaller success rate of the unbalanced protocol, followed by the same key reduction during privacy amplification as for a balanced protocol. So in effect, during the operation of an unbalanced protocol, the use of privacy amplification terms from a balanced BB84 protocol still gives valid secret key rates. Therefore, it is incorrect for Ref. \cite{Agnes} to conclude that the drop in secure key rates for the unbalanced cases is due to the application of a new security analysis. Since Ref. \cite{Kiyoshi} provides a known analytical key rate of this scenario, we use that result as the baseline of our investigations to show that in fact the secret key rate is underestimated by this security analysis, and thus less privacy amplification is required in this situation.

\begin{figure}[!htb]
    \centering
    \includegraphics[width=1.1\linewidth]{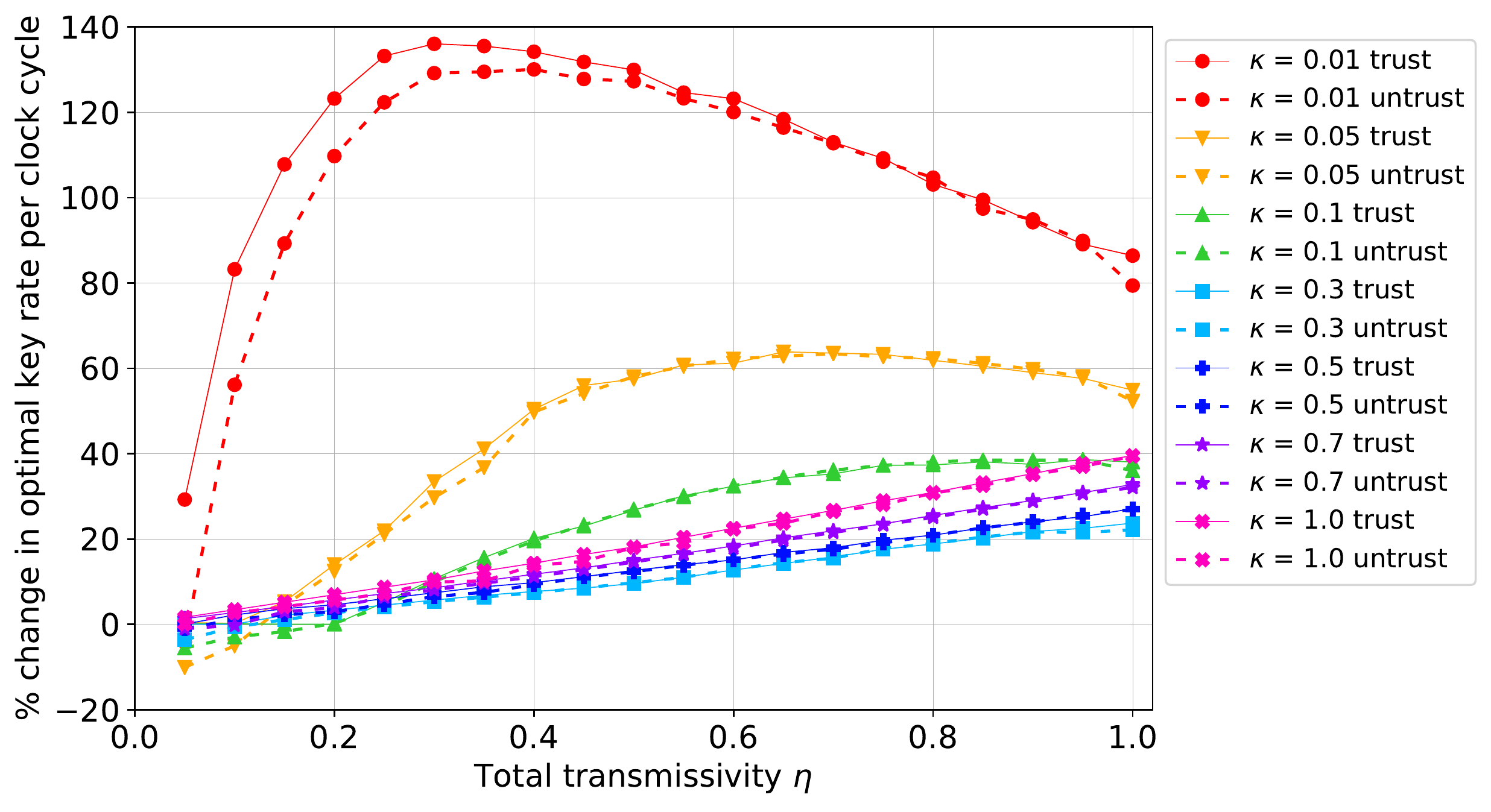}
    \caption{Percentage change in key rates comparing our optimal lower bounds for key rates with \cite{Kiyoshi}'s optimal key rates versus total transmissivity $\eta$. We label the changes for trusted (untrusted) dark counts with solid (dotted) lines. A positive change means that our key rate is higher.\label{CompareKiyoshi}}
\end{figure}

We compare our key rates with \cite{Kiyoshi}'s in Fig.\;\ref{CompareKiyoshi}, which shows that our analysis provides higher key rates for total transmissivity $\eta>0.1$ ($<$10 dB), especially for small $\kappa$ values. Our method shows advantage in low-loss cases because the PA components from the multi-photon part of Alice's signals are larger in the low-loss regime, which are pessimistically set to zero in \cite{Kiyoshi}. This can be understood as Eve does not learn too much of the multi-photon signals, thereby allowing more information to reach Bob.

When the total transmissivity satisfies $\eta\leq0.2$, we encounter the issue with infeasible constraints with untrusted dark counts. We recover approximately the same key rates in \cite{Kiyoshi} for most cases, but some of our lower bounds for the key rates (obtained from maximising the dual SDP problem) in the untrusted noise scenario appear to be slightly lower than \cite{Kiyoshi}'s. To understand the gaps between our key rate upper bounds (which are on par with \cite{Kiyoshi}'s key rates) and lower bounds (see Sec.\;\ref{NumericalImplementation} for the meaning of the two bounds), we recall that our way of getting around the infeasibility issue with untrusted noise is to relax the required precision for the constraints to be satisfied in the numerical solver. The first-step suboptimal solution to the relaxed problem will naturally suffer from stronger constraint violations which lead to a larger penalty term in the calculation of the dual suboptimal solution \cite{AdamNumerics}.

Notice that when the asymmetric loss parameter reaches $\kappa = 0.3$, the percentage increase of our key rate relative to \cite{Kiyoshi}'s is the least compared to other values of $\kappa$. This phenomenon is also observed when we make the following choices of parameters: flag-state photon cutoff $N_B\in\{1,2,3,4\}$, dark count probability $p_d\in\{0, 10^{-5}, 10^{-4}\}$, and total transmissivity $\eta = 1$. As our numerical data suggest, the ratio between the optimal values of the privacy amplification terms attributed to Alice sending out 1-photon and 2-photon signals,
\begin{equation}
    r_{21} = \frac{\min_{\rho_{AB}^2\in\mathbf{S}_2}D(\mathcal{G}(\rho_{AB}^2)||\mathcal{Z}(\mathcal{G}(\rho_{AB}^2)))}{\min_{\rho_{AB}^1\in\mathbf{S}_1}D(\mathcal{G}(\rho_{AB}^1)||\mathcal{Z}(\mathcal{G}(\rho_{AB}^1)))},
\end{equation}
reaches its smallest value when $\kappa \approx 0.3$. This can be interpreted as the amount of private information carried by 2-photon signals relative to the amount carried by 1-photon signals is the least when $\kappa\approx 0.3$, which corresponds to the points with the least key rate improvement. 

As a remark, the optimal signal intensities $|\widetilde{\alpha}_\text{opt}|^2$'s for \cite{Kiyoshi}'s optimal key rates (corresponding to Eqn. (6) in \cite{Kiyoshi}), which we compare with in Fig.\;\ref{CompareKiyoshi}, are slowly decreasing as $\eta$ increases. They satisfy $|\widetilde{\alpha}_\text{opt}|^2\leq\min\{1,|\alpha_\text{opt}|^2\}$ where $|\alpha_\text{opt}|^2$ is the corresponding optimal intensity of our analysis as plotted in Fig. \ref{OptimalKeyRate}(b). This means that \cite{Kiyoshi}'s optimal signal intensity is always smaller than our optimal intensity $|\alpha_\text{opt}|^2$. It is also true that \cite{Kiyoshi}'s optimal intensity increases as $\kappa$ reduces for all tested values of $\eta$.

\begin{figure}[hbt!]
    \centering
    \includegraphics[width=1.1\linewidth]{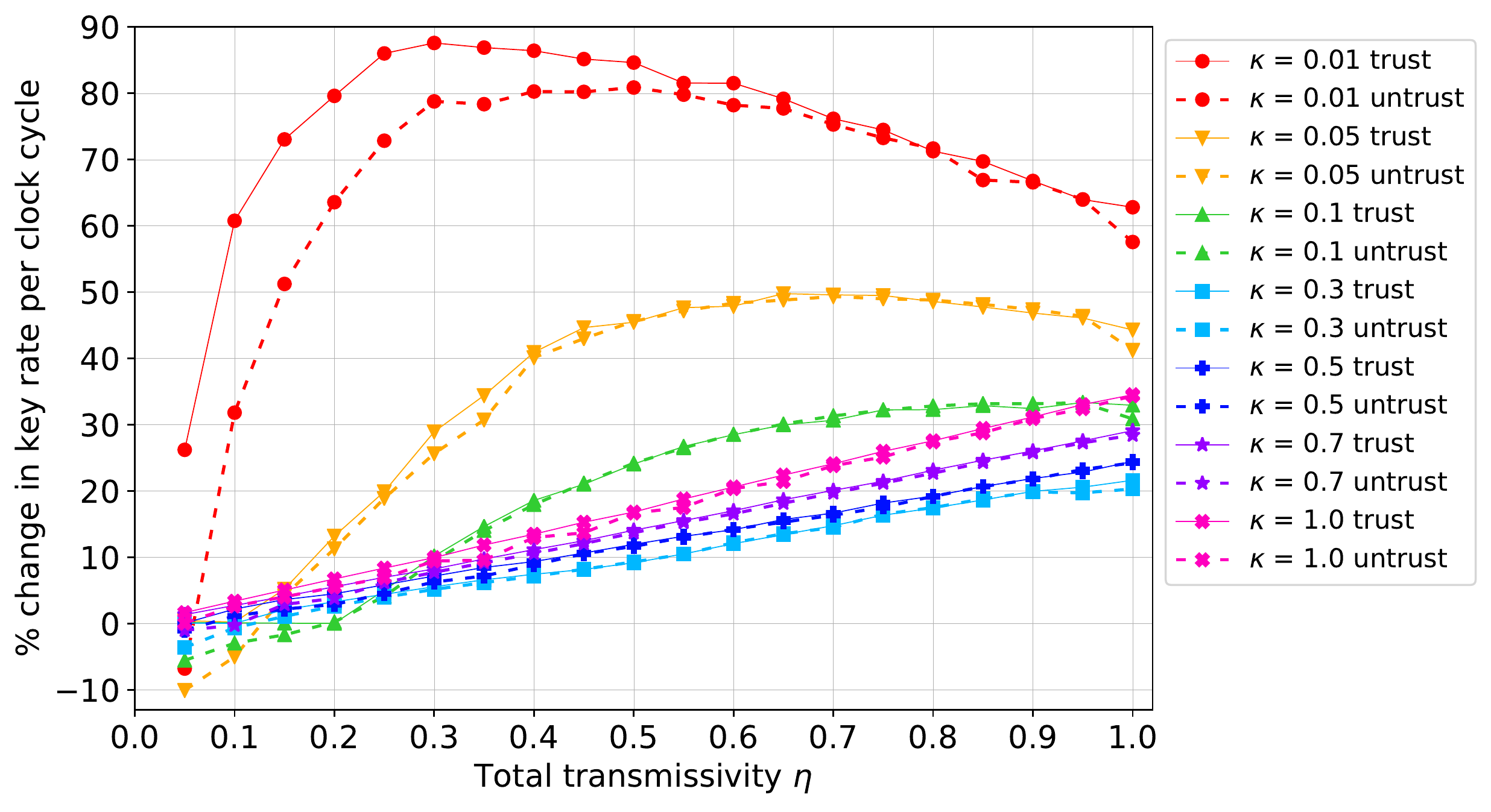}
    \caption{Percentage change in key rates comparing our lower bounds for key rates per clock cycle evaluated at \cite{Kiyoshi}'s optimal $\widetilde{\alpha}_\text{opt}$ with \cite{Kiyoshi}'s optimal key rates versus total transmissivity $\eta$. We label the changes for trusted (untrusted) dark counts with solid (dotted) lines. A positive change means that our key rate is higher.
    \label{KeyRate-Kiyoshi_WithOptKiyoshiAlpha_PercentChange}}
\end{figure}

In the post-processing view, the goal is to determine the amount of key reduction from privacy amplification that guarantees a secure final key for a given set of experimental parameters. Particularly, in the case where the attenuation of the laser has already been set to \cite{Kiyoshi}'s optimal intensity for a chosen set of parameters, we compare the privacy amplification term from our analysis with the one from \cite{Kiyoshi}'s. To see this, we first show in Fig.\;\ref{KeyRate-Kiyoshi_WithOptKiyoshiAlpha_PercentChange} that our method still gives higher key rates than \cite{Kiyoshi}'s in the low-loss regime ($\eta>0.15$) even when our signal intensities are set to \cite{Kiyoshi}'s. We then make the connection between this result and the difference in privacy amplification with two observations: 1) the probability of passing post-selection $p_{\text{pass}}$ is equal for both methods and 2) the costs of error correction are approximately equal when the same signal intensity is used in both approaches. It follows that the difference in key rates translates to the difference in the privacy amplification terms in the key rate formula. Thus, our method requires less key reduction from privacy amplification compared to \cite{Kiyoshi} for low-loss scenarios. This allows us to extract more secret key out of these unbalanced protocols than previously thought.

We now turn to study the effect of trusted loss on the key rates. Previously, we assume that the quantum channel contributes completely to the total loss. However, if we know that a certain part of the total loss is caused by some trusted components (e.g. Bob's detectors), the key rate can be improved since the channel loss is effectively smaller. The key rate improvement has already been shown in both active and passive BB84 protocol \cite{Yanbao} where the detection efficiency of the receiver's detectors is assumed to be beyond Eve's control. We will present a similar behaviour of the key rates of this protocol under different trusted loss conditions.

\begin{figure}[!htb]
\renewcommand{\arraystretch}{0}
\centering
\begin{tabular}{@{}c@{}}
    \includegraphics[width=0.95\linewidth]{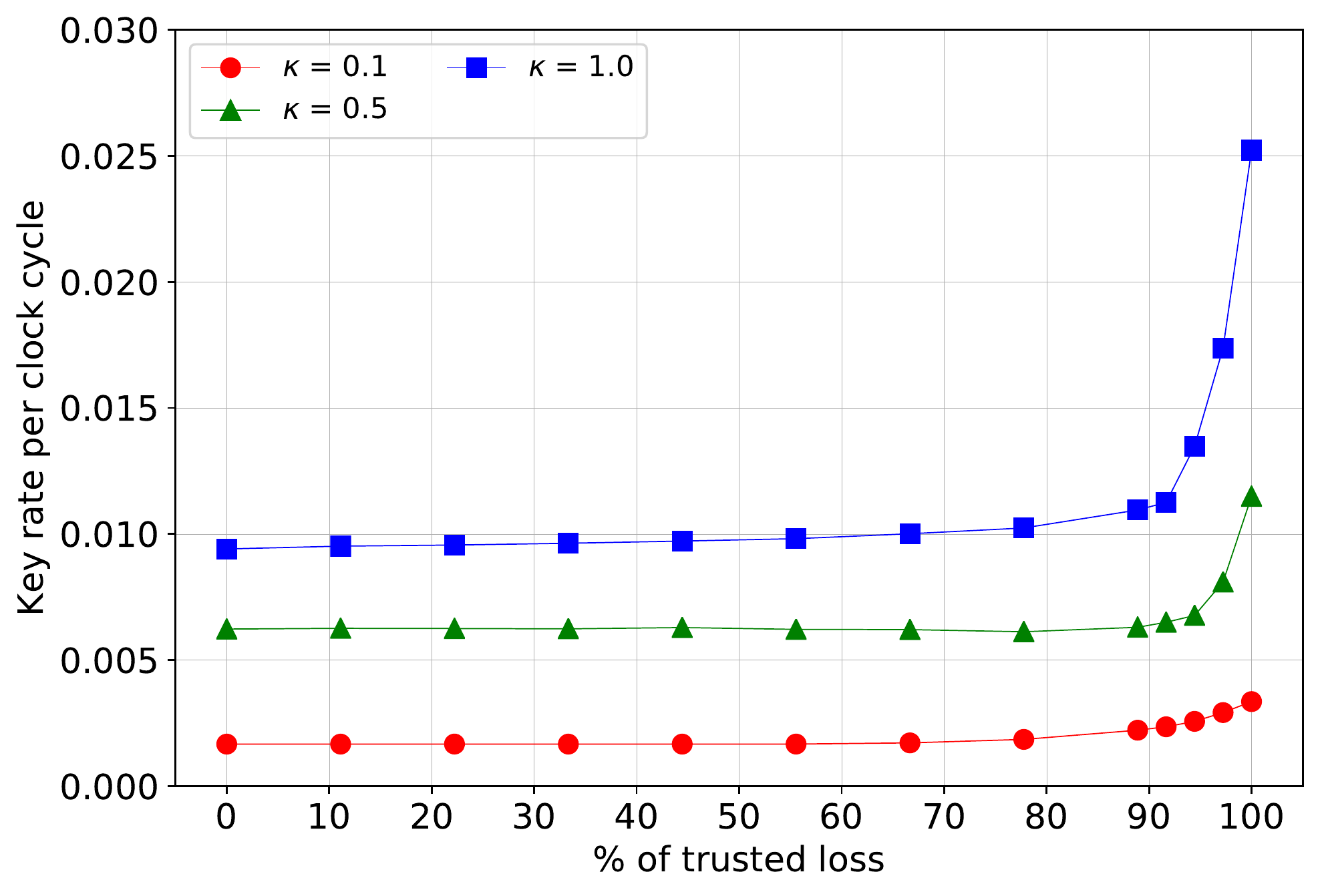}\\
    \text{(a)}\\
    \includegraphics[width=0.95\linewidth]{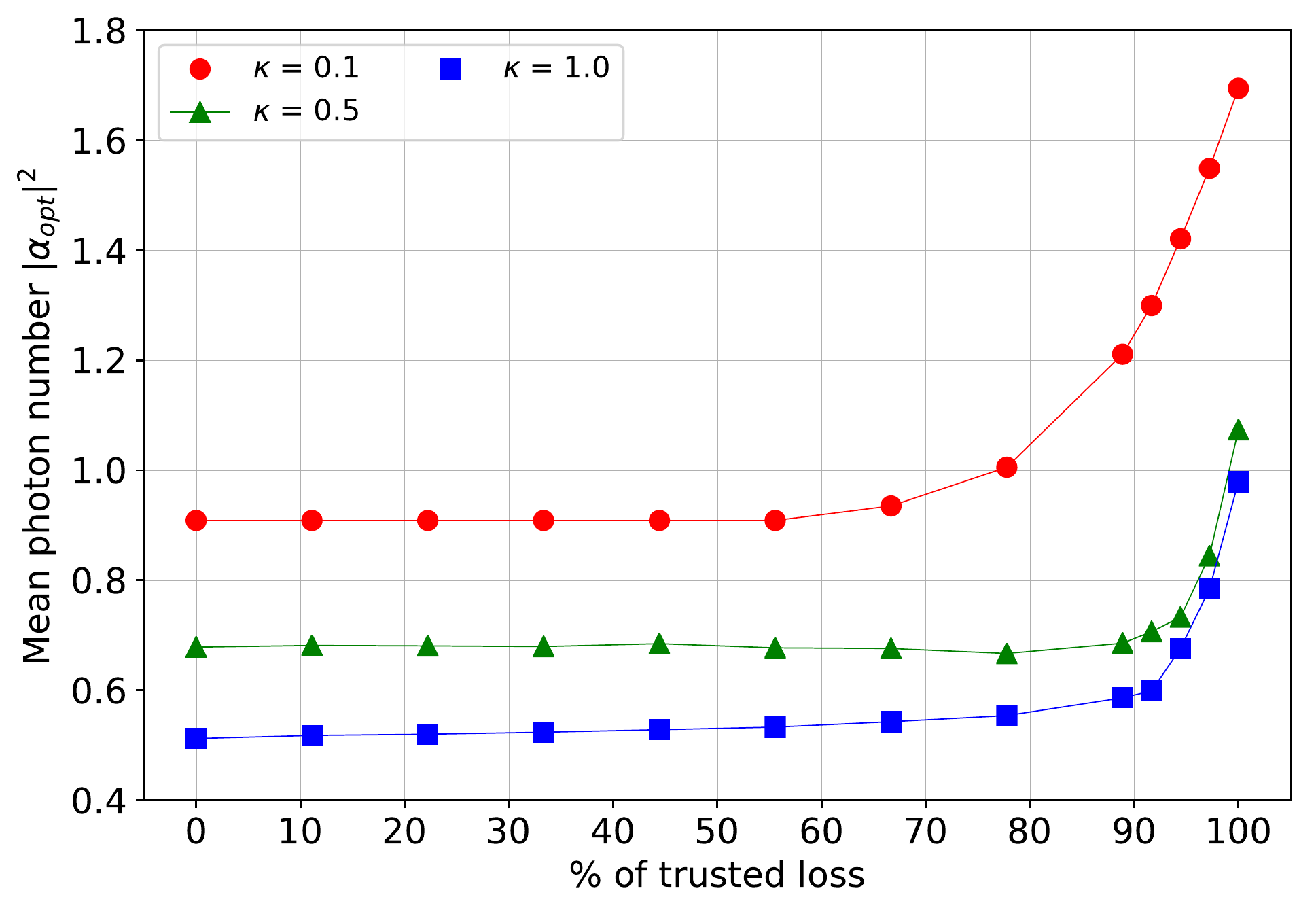} \\
    \text{(b)}
  \end{tabular}
\caption{Assuming trusted dark counts, (a) our lower bounds for key rates and (b) the mean photon numbers plotted against the proportion (in percentage) of the trusted loss coming from the detection inefficiency of Bob's detectors to a fixed total loss corresponding to total transmissivity $\eta=0.1$.}\label{KeyRate_DetIneff}
\end{figure}

We fix the total transmissivity to be $\eta=0.1$ and assume dark counts to be trusted, and then we vary the detection efficiency of Bob's trusted detectors $\eta_\text{det}$. Indeed, Fig.\;\ref{KeyRate_DetIneff}(a) shows that the lower bound of our optimal key rate increases with the proportion of the trusted loss component coming from Bob's detectors to the total loss, which takes the form $\frac{1-\eta_\text{det}}{1-\eta}$. The optimal mean photon numbers corresponding to the optimal key rates are displayed in Fig. \ref{KeyRate_DetIneff}(b).

To summarise this section, we report a significant gain in key rates in the low-loss regime ($<$10 dB) with our analysis. To be precise, with our security analysis, higher key rates can be obtained when the signal intensities are set to our optimal and \cite{Kiyoshi}'s optimal values. We emphasise that the reported improvement can be attained without any modification to the experimental setup. Lastly, we show that the key rates can be increased if we know that the detection inefficiency contributes a considerable amount to the total loss.

\section{Summary \& Outlook}\label{Conclusion}
This work provides a new numerical security proof for the unbalanced phase-encoded BB84 protocol. Using the newly developed flag-state squashing model \cite{Yanbao}, we are able to derive additional private information from the multi-photon components of the signal states. We compare our key rates with the key rates proved in Ref. \cite{Kiyoshi} under the same simulation parameters and show that our analysis results in significantly higher key rates in the low-loss regime. In the design view, we find that a balanced protocol ($\kappa =1$) gives a higher key rate than an unbalanced protocol so that a design cannot take advantage of an artificial induction of asymmetry. In the post-processing view, our method requires less key reduction from privacy amplification compared to \cite{Kiyoshi} for low-loss cases. We prove that our key rates are still better than \cite{Kiyoshi}'s even when their optimal mean signal photon numbers are used. Hence, any experiments that are already implementing the optimal settings of \cite{Kiyoshi} can profit from our higher key rates. We also explore the advantage of characterising the receiver's detection inefficiency as a trusted loss, which is not allowed by \cite{Agnes,Kiyoshi}'s proof technique. Our results suggest that the key rate can be improved when the proportion of trusted loss due to detection inefficiency to the total loss is significant.

Let us conclude by pointing out some future directions of investigation: It is important to find a formal way of incorporating untrusted dark counts into the security analysis without leading to unphysical constraints. As mentioned in Sec. \ref{Sec:decoy}, to extend our analysis to the use of a finite number of decoy states, one must consider the dependence among different feasible conditional state sets when handling the privacy amplification term. Finally, some of our proof techniques can be transferred to a finite-key analysis. It would be worth comparing the key rates from a finite-key analysis \cite{IanFiniteSizePaper} with the asymptotic key rates reported here.

\begin{acknowledgments}
The first author would like to thank Jie Lin and Shlok Nahar for reviewing his MATLAB code which produces the results in this paper. We would also like to thank Ian George and Jie Lin for proofreading the appendices, and Adam Winick for the useful discussion on the numerical simulation of experimental statistics. The work has been performed at the Institute for Quantum Computing, University of Waterloo, which is supported by Industry Canada. The research has been supported by NSERC under the Discovery Grants Program, Grant No. 341495, and under the Collaborative Research and Development Program, Grant No. CRDP J 522308-17. Financial support for this work has been partially provided by Huawei Technologies Canada Co., Ltd. This research was enabled in part by the computational resources provided by WestGrid (www.westgrid.ca) and Compute Canada (www.computecanada.ca).

\end{acknowledgments}

\appendix

\section{Derivation of the lower bound for the weight of $(n\leq N_B)$-photon signal subspace} \label{App:LowerBound}

We aim at lower bounding the weight of the $(n\leq\;N_B)$-photon signal subspace, $p(n$\hspace{0pt}$\leq$\hspace{0pt}$N_B)$, with Bob's observed statistics. In this appendix, we use the cross-click probability to derive a lower bound for $p(n$\hspace{0pt}$\leq$\hspace{0pt}$N_B)$ in the following steps. The cross-click probability for any signal satisfies
\begin{flalign}
    p(\text{cc}) &= \sum^{N_B}_{n=0}p(n)p(\text{cc}|n)+\sum^\infty_{n=N_B+1}p(n)p(\text{cc}|n) \nonumber\\
    &\geq \sum^{N_B}_{n=0}p(n)p_{\text{min}}(\text{cc}|n)+\sum^\infty_{n=N_B+1}p(n)p_{\text{min}}(\text{cc}|n) \nonumber\\
    &\geq p(n\leq N_B) C^{\text{min}}_{n\leq N_B} + [1-p(n\leq N_B)] C^{\text{min}}_{n>N_B} \nonumber\\
    &= C^{\text{min}}_{n>N_B} - p(n\leq N_B)(C^{\text{min}}_{n>N_B} - C^{\text{min}}_{n\leq N_B}). \label{PccIneqDerivation}
\end{flalign}
In the second line, $p_{\text{min}}(\text{cc}|n)$ denotes the minimal cross-click probability given that Bob receives an $n$-photon signal. In the last two lines, we define $p(n$\hspace{0pt}$\leq$\hspace{0pt}$N_B) \coloneqq \sum^{N_B}_{n=0}p(n)$, $C^{\text{min}}_{n\leq N_B} \coloneqq \min_{0\leq n\leq N_B}p_{\text{min}}(\text{cc}|n)$ and $C^{\text{min}}_{n>N_B} \coloneqq \min_{n>N_B}p_{\text{min}}(\text{cc}|n)$. If $p_{\text{min}}(\text{cc}|n)$ is monotonically increasing with $n$, then $C^{\text{min}}_{n\leq N_B} = p_{\text{min}}(\text{cc}|0)$ and $C^{\text{min}}_{n>N_B} = p_{\text{min}}(\text{cc}|N_B+1)$. If we also have strict inequality $C^{\text{min}}_{n>N_B} > C^{\text{min}}_{n\leq N_B}$, then we can turn the inequality in (\ref{PccIneqDerivation}) into the desired lower bound 
\begin{equation}
    p(n\leq N_B) \geq 1 - \frac{p(\text{cc}) - p_{\text{min}}(\text{cc}|0)}{p_{\text{min}}(\text{cc}|N_B+1) - p_{\text{min}}(\text{cc}|0)} \eqqcolon B^\text{min}_{n\leq N_B}\;. \label{PleqN_B_Derivation}
\end{equation}
We will show that the minimum cross-click probabilities indeed satisfy the monotonicity and the strict inequality conditions.

To obtain the minimum conditional probabilities $p_{\text{min}}(\text{cc}|0)$ and $p_{\text{min}}(\text{cc}|N_B+1)$, we start by considering the new POVM elements after classical post-processing due to dark counts as mentioned in Sec. \ref{POVMSec} which are
\begin{flalign}
    P_{0}^{\phi_B} &= (1-p_d)^6 \;F_{0}^{\phi_B}\;, \label{Eq:CPP0}\\
    P_{t_1}^{\phi_B} &= (1-p_d)^4(F_{t_1}^{\phi_B} + (1-(1-p_d)^2)F_{0}^{\phi_B}),\\
    P_{t_3}^{\phi_B} &= (1-p_d)^4(F_{t_3}^{\phi_B} + (1-(1-p_d)^2)F_{0}^{\phi_B}),\\
    P_{2}^{\phi_B} &= (1-p_d)^5(F_{2}^{\phi_B} + p_d\;F_{0}^{\phi_B}),\\
    P_{5}^{\phi_B} &= (1-p_d)^5(F_{5}^{\phi_B} + p_d\;F_{0}^{\phi_B}),\\
    P_{t_1,t_3}^{\phi_B} &= (1-p_d)^2\{F_{t_1,t_3}^{\phi_B} + [1-(1-p_d)^2](F_{t_1}^{\phi_B}+F_{t_3}^{\phi_B})\nonumber\\ 
    &\text{\hspace{54pt}} + [1-(1-p_d)^2]^2\;F_{0}^{\phi_B}\}, \\
    P_{2,5}^{\phi_B} &= (1-p_d)^4[F_{2,5}^{\phi_B} + p_d(F_{2}^{\phi_B}+F_{5}^{\phi_B}) + p_d^2\;F_{0}^{\phi_B}]. \label{Eq:CPP25}
\end{flalign}
We first group the pre-processed POVM elements into two coarse-grained POVM elements: outside-only ($t_1$, $t_3$, $t_1$\&$t_3$) and inside-only (2, 5, 2\&5). Using Eqns.\;(\ref{Eq:singleclick}) and (\ref{Eq:doubleclick}), the two elements can be expressed as
\begin{flalign}
    F_{\text{out}} &= \sum_{\phi_B \in\{0,\pi/2\}}(F_{t_1,t_3}^{\phi_B} +F_{t_1}^{\phi_B}+F_{t_3}^{\phi_B}) \nonumber\\
    &= \sum_{n=1}^\infty\sum_{i=0}^n \xi^i(1-\xi)^{n-i}|i,n-i\rangle\langle i, n-i|, \\
    F_{\text{in}} &= \sum_{\phi_B \in\{0,\pi/2\}}(F_{2,5}^{\phi_B} +F_{2}^{\phi_B}+F_{5}^{\phi_B}) \nonumber \\
    &= \sum_{n=1}^\infty\sum_{i=0}^n \xi^{n-i}(1-\xi)^i|i,n-i\rangle\langle i, n-i| .
\end{flalign}
Similarly, the two coarse-grained post-processed POVM elements can be found to be
\begin{flalign}
    P_{\text{out}} &= \sum_{\phi_B \in\{0,\pi/2\}}(P_{t_1,t_3}^{\phi_B} +P_{t_1}^{\phi_B}+P_{t_3}^{\phi_B})  \nonumber\\
    &= (1-p_d)^2 \{F_{\text{out}} + [1-(1-p_d)^4]F_{0}\},\\
    P_{\text{in}} &= \sum_{\phi_B \in\{0,\pi/2\}}(P_{2,5}^{\phi_B} +P_{2}^{\phi_B}+P_{5}^{\phi_B})  \nonumber\\
    &= (1-p_d)^4 \{F_{\text{in}} + p_d(2-p_d)F_{0}\},
\end{flalign}
where the pre-processed no-click POVM element is $F_0 = |0,0\rangle\langle 0,0|$. Therefore, the post-processed coarse-grained POVM elements for inside-only and outside-only clicks are diagonal in the two-mode Fock basis $\{|i,n-i\rangle: i=0,...,n\}$ for all $n\in\mathbb{N}$. The cross-click POVM element is 
\begin{equation}
    P_{\text{cc}} = \mathbb{1}_B - (P_{\text{out}} + P_{\text{in}} + \sum_{\phi_B} P_{0}^{\phi_B}) \label{Eq:Pcc}
\end{equation}
which is also diagonal in the two-mode Fock basis. Since $P_{\text{cc}}$ is already diagonal, it is straightforward to find $P_{\text{cc}}$'s minimum eigenvalue restricted to the $n$-photon subspace, which corresponds to the minimum cross-click probability for any $n$-photon input states, analytically. For an eigenstate $|i,n-i\rangle$, the associated cross-click probability (the eigenvalue of $P_{\text{cc}}$) can be found using Eqns.\;(\ref{Eq:CPP0})\;--\;(\ref{Eq:Pcc}) as 
\begin{flalign}
    p(\text{cc}|\;|i,n-i\rangle) &= 1 - (1-p_d)^2\;\xi^i(1-\xi)^{n-i}  \nonumber\\
    &\text{\hspace{19pt}} - (1-p_d)^4\;\xi^{n-i}(1-\xi)^i
\end{flalign}
for $n\geq 1$, and for the vacuum state $|0,0\rangle$ to be
\begin{equation}
    p(\text{cc}|0) = 1 - (1-p_d)^2 [1 + p_d\; (1-p_d)^2 (2-p_d)] \nonumber
\end{equation}
as stated in Eqn.\;(\ref{pcc|0}). Since there is only one eigenvalue in the vacuum subspace, we need not minimise the conditional probability (i.e. $p_{\text{min}}(\text{cc}|0) = p(\text{cc}|0)$).
We exclude the case where the phase modulator has zero transmissivity ($\kappa=0$), then $\xi = \frac{1}{1+\kappa}\in [\frac{1}{2},1)$, so the minimum cross-click probability for any $(n\geq1)$-photon input state is 
\begin{equation}
    p_{\text{min}}(\text{cc}|n) = 1 - (1-p_d)^2\; \xi^{n} - (1-p_d)^4\; (1-\xi)^{n} \nonumber
\end{equation}
as stated in Eqn.\;(\ref{p_min_cc|n}), which is valid for all $n\geq1$. Notice that $p_{\text{min}}(\text{cc}|n)$ is monotonically increasing with $n$ which agrees with our intuition that cross-click events are more likely with more incoming photons.

As we further restrict the dark count probability to $p_d \in [0,1)$, it is analytically straightforward to verify that for all $n\geq1$ and $\xi \in [\frac{1}{2},1)$,
\begin{equation}
    p(\text{cc}|0) \leq p_{\text{min}}(\text{cc}|n) < p_{\text{min}}(\text{cc}|n+1),
\end{equation}
so the monotonicity and the strict inequality conditions for (\ref{PleqN_B_Derivation}) to hold are satisfied. The inequality (\ref{PleqN_B_Derivation}) is of the same form as (\ref{lowerbound}) in Sec. \ref{FlagStateSection} except that the observed cross-click probability in (\ref{lowerbound}) is conditioned on Alice's signal choice $x$.

We now move on to prove that the lower bound in the inequality (\ref{PleqN_B_Derivation}) is tighter than the lower bound derived in Ref. \cite{VarunThesis} for no dark counts. We use the fact that
\begin{equation}
    \frac{a-c}{b-c} \leq \frac{a}{b} \text{\hspace{8pt}, if \hspace{1pt}} 0\leq c \leq a\leq b
\end{equation}
and all probabilities are positive to show that
\begin{equation}
    \frac{p(\text{cc}) - p(\text{cc}|0)}{p_{\text{min}}(\text{cc}|N_B+1) - p(\text{cc}|0)} \leq \frac{p(\text{cc})}{p_{\text{min}}(\text{cc}|N_B+1)}\;.
\end{equation}
With (\ref{p_min_cc|n}), we can further show that 
\begin{equation}
    p_{\text{min}}(\text{cc}|N_B+1) \geq 1-\xi^{N_B+1}-(1-\xi)^{N_B+1}\;.
\end{equation}
Thus, the lower bound in (\ref{PleqN_B_Derivation}) is larger than the lower bound derived in Ref. \cite{VarunThesis} which is the expression in (\ref{p_min_cc|n}) for zero dark-count rate as in
\begin{equation}
    p(n\leq N_B) \geq B^\text{min}_{n\leq N_B} \geq 1 - \frac{p(\text{cc})}{1-\xi^{N_B+1}-(1-\xi)^{N_B+1}}\;.
\end{equation}
The secure key rate should only reduce as we loosen the lower bound for the $(n$\hspace{0pt}$\leq$\hspace{0pt}$N_B)$-photon subspace since the flag-state squashing map can be more entanglement-breaking and so Eve could gain more information from purification. As a result, we can use the dark-count-free lower bound blindly on Bob's measurement data to obtain a secure key rate even if the dark-count rate is assumed to be zero.

\section{Proof of Decomposing the Privacy Amplification term}\label{PADecompApp}

In Sec.\;\ref{StatePrep}, Eqns.\;(\ref{snx}), (\ref{sigmax}) and (\ref{AliceFullState}) together describe the entangled pure state that Alice prepares to be
\begin{equation}
    |\Psi\rangle_{AA_SA'} = \sum_x \sqrt{p_x}\;|x\rangle_A\otimes\sum_{\widetilde{n}=0}^\infty \sqrt{p_{\widetilde{n}}}\;|\widetilde{n}\rangle_{A_S}\otimes|s^x_{\widetilde{n}}\rangle_{A'}
    \nonumber
\end{equation}
where we simplify the notation here with $p_{\widetilde{n}} \coloneqq p_{\widetilde{n}}(\frac{\alpha}{\sqrt{\xi}})$. Since the phase-randomised coherent signal states are block-diagonal in total photon number basis in Eve's point of view, Eve can, without loss of generality, perform QND measurements to determine the total photon number in the signal states. This allows her to keep an extra classical register that tells her the
total number of photons in the signal without degrading her eavesdropping power as we will see below.

To see why allowing Eve to measure the total photon number in the signal state will not affect our security statement, we first consider the most general scenario where we do not assume anything about Eve's attack. By Stinespring's dilation theorem, the action of a quantum channel on the signal state can be described by an isometry $V_{A'\rightarrow BE}$ that takes Alice's signal system, $A'$, to Bob's system, $B$, and Eve's purifying system, $E$, such that the pure state shared among all parties is
\begin{equation}
    |\widetilde{\Psi}\rangle_{AA_SBE} = \sum_x \sqrt{p_x}\;|x\rangle_A\otimes\sum_{\widetilde{n}=0}^\infty \sqrt{p_{\widetilde{n}}}\;|\widetilde{n}\rangle_{A_S}\otimes V_{A'\rightarrow BE}\;|s^x_{\widetilde{n}}\rangle_{A'}\;.\nonumber
\end{equation}
Eve's general reduced state conditioned on Alice's measurement outcome $x$ is
\begin{equation}
    \rho_{E}^x = \sum_{\widetilde{n}=0}^\infty p_{\widetilde{n}}\; \Tr_B(V_{A'\rightarrow BE}\;|s^x_{\widetilde{n}}\rangle\langle s^x_{\widetilde{n}}|\;V_{A'\rightarrow BE}^\dagger). \label{rhoExGeneral}
\end{equation}

In the alternative scenario, we assume that Eve performs the QND measurement and could perform adaptive attack according to her knowledge of the photon number. Let Eve's purifying system of the signal be $E$ and the extra register for recording the photon number in Alice's signal be $\widetilde{E}$. Again by Stinespring's dilation theorem, one can describe the action of a quantum channel on the signal state by an isometry $V_{A'\rightarrow BE\widetilde{E}}$ which takes the form
\begin{equation}
    V_{A'\rightarrow BE\widetilde{E}} = \sum_{\widetilde{n}=0}^\infty V_{A'\rightarrow BE}^{\widetilde{n}}\;\Pi_{\widetilde{n}}^{A'}\otimes |\widetilde{n}\rangle_{\widetilde{E}}
\end{equation}
where $V_{A'\rightarrow BE}^{\widetilde{n}}$ is Eve's isometry for purifying Bob's quantum state given that she learns the total photon number $\widetilde{n}$ and $\Pi_{\widetilde{n}}^{A'}$ is a projector which projects onto the $\widetilde{n}$-total photon subspace of the signal system $A'$. The shared pure state between Alice, Bob and Eve before any announcements is
\begin{widetext}
\begin{equation}
    |\Psi\rangle_{AA_SBE\widetilde{E}} = \sum_x \sqrt{p_x}\;|x\rangle_A\otimes\sum_{\widetilde{n}=0}^\infty \sqrt{p_{\widetilde{n}}}\;|\widetilde{n}\rangle_{A_S}\otimes V_{A'\rightarrow BE}^{\widetilde{n}}\;|s^x_{\widetilde{n}}\rangle_{A'}\otimes|\widetilde{n}\rangle_{\widetilde{E}}
 \label{eq:wideeq}
\end{equation}
\end{widetext}
and Eve's reduced state conditioned on Alice's measurement outcome $x$ is
\begin{equation}
    \rho_{E\widetilde{E}}^x = \sum_{\widetilde{n}=0}^\infty p_{\widetilde{n}}\; \Tr_B[V_{A'\rightarrow BE}^{\widetilde{n}}\;|s^x_{\widetilde{n}}\rangle\langle s^x_{\widetilde{n}}|\;(V_{A'\rightarrow BE}^{\widetilde{n}})^\dagger]\otimes|\widetilde{n}\rangle\langle \widetilde{n}|_{\widetilde{E}}\;.
\end{equation}
If we further trace out Eve's register $\widetilde{E}$, her reduced state $\rho_{E}^x$ clearly contains the general attack in (\ref{rhoExGeneral}) where Eve performs the same purification (i.e. $V_{A'\rightarrow BE}^{\widetilde{n}} = V_{A'\rightarrow BE}$) for all $\widetilde{n}\in\mathbb{N}$. Therefore, the assumption that Eve can measure the photon number of the signal and the pure state shared by all parties to be (\ref{eq:wideeq}) will not affect the security statement of our proof.

To decompose the relative entropy in Eqn.\;(\ref{KeyRateFormula2}), we can assume the pure state shared by all parties to be (\ref{eq:wideeq}) as argued above. Hence, the state shared by Alice and Bob is
\begin{equation}
    \rho_{AA_SB} = \sum_{x,y} \sqrt{p_x p_y}\;|x\rangle\langle y|_A\otimes\sum_{\widetilde{n}=0}^\infty p_{\widetilde{n}}\;|\widetilde{n}\rangle\langle \widetilde{n}|_{A_S}\otimes \Phi(|s^x_{\widetilde{n}}\rangle\langle s^y_{\widetilde{n}}|), \label{rhoAASB_QND}
\end{equation}
where the quantum channel between Alice and Bob is defined as $\Phi(\cdot)\coloneqq \Tr_E(V_{A'\rightarrow BE}\cdot V_{A'\rightarrow BE}^\dagger)$. If we reorder the positions of the three registers in the tensor product and define the conditional state $\rho^{\widetilde{n}}_{AB} = \sum_{x,y} \sqrt{p_x p_y}\;|x\rangle\langle y|_A\otimes \Phi(|s^x_{\widetilde{n}}\rangle\langle s^y_{\widetilde{n}}|)$, the state in (\ref{rhoAASB_QND}) can be expressed as
\begin{equation}
    \rho_{AA_SB} = \sum_{\widetilde{n}=0}^\infty p_{\widetilde{n}}\;|\widetilde{n}\rangle\langle \widetilde{n}|_{A_S}\otimes \rho^{\widetilde{n}}_{AB}\;. \label{rhoAsAB_reorder}
\end{equation}
We will utilise this block-diagonal structure to decompose the relative entropy $D(\mathcal{G}(\rho_{AA_SB})||\mathcal{Z}(\mathcal{G}(\rho_{AA_SB})))$ in the following steps.

According to the definitions of $\mathcal{G}$ and $\mathcal{Z}$ maps stated in Eqns.\;(\ref{Gmap})\;--\;(\ref{Zmap}), both maps act trivially on Alice's shield system $A_S$ (i.e. apply $\mathbb{1}_{A_S}$ to the input state). Hence, the unnormalised states $\mathcal{G}(\rho_{AA_SB})$ and $\mathcal{Z}(\mathcal{G}(\rho_{AA_SB}))$ are also block-diagonal as in
\begin{equation}
    \mathcal{N}(\rho_{AA_SB}) = \sum_{\widetilde{n}=0}^\infty p_{\widetilde{n}}\;|\widetilde{n}\rangle\langle \widetilde{n}|_{A_S}\otimes \mathcal{N}(\rho^{\widetilde{n}}_{AB})
\end{equation}
for $\mathcal{N}$ to be the substitute for the maps $\mathcal{G}$ and $\mathcal{Z}\circ\mathcal{G}$. Taking the matrix logarithm gives us
\begin{equation}
    \log \mathcal{N}(\rho_{AA_SB}) = \sum_{\widetilde{n}=0}^\infty |\widetilde{n}\rangle\langle \widetilde{n}|_{A_S}\otimes [(\log p_{\widetilde{n}}) \mathbb{1} + \log\mathcal{N}(\rho^{\widetilde{n}}_{AB})].
\end{equation}
By the definition of relative entropy, we decompose the PA term into
\begin{flalign}
    & \hspace{1pt}D(\mathcal{G}(\rho_{AA_SB})||\mathcal{Z}(\mathcal{G}(\rho_{AA_SB}))) \nonumber\\
    = & \hspace{1pt}\Tr\{\mathcal{G}(\rho_{AA_SB})\left[\log \mathcal{G}(\rho_{AA_SB}) - \log \mathcal{Z}(\mathcal{G}(\rho_{AA_SB}))\right]\} \nonumber\\
    = & \hspace{1pt}\sum_{\widetilde{n}=0}^\infty p_{\widetilde{n}}\; \Tr\{\mathcal{G}(\rho_{AB}^{\widetilde{n}})\left[\log \mathcal{G}(\rho_{AB}^{\widetilde{n}}) - \log \mathcal{Z}(\mathcal{G}(\rho_{AB}^{\widetilde{n}})\right]\} \nonumber\\
    = & \hspace{1pt}\sum_{\widetilde{n}=0}^\infty p_{\widetilde{n}}\; D(\mathcal{G}(\rho_{AB}^{\widetilde{n}})||\mathcal{Z}(\mathcal{G}(\rho_{AB}^{\widetilde{n}}))),
\end{flalign}
which completes the proof.

\section{Justifications for speeding up numerical optimisations}\label{App:SpeedupMain}
\subsection{Reducing the number of variables}\label{App:Speedup1}

To speed up the optimisation for the problem specified in (\ref{OPT}), we make use of the structure of the flag-state squashed state. The joint state shared between Alice and Bob $\rho_{AB}$ can be expressed as
\begin{equation}
    \rho_{AB} = \sum_{i,j=1}^{d_A}\sum_{n,m=1}^\infty \rho_{i,j}^{n,m} E_{i,j}\otimes E_{n,m}\;\;,
\end{equation}
where $E_{i,j}=|i\rangle\langle j|$ with $\{|i\rangle\}$ being an orthonormal basis and $\rho_{i,j}^{n,m}\in\mathbb{C}$ $\forall$ $i,j,n,m$. Recall that the flag-state squashing map takes the form of (\ref{flagstateSquash}) and since the dimension of the 2-mode $(n\leq N_B)$-photon subspace is $\Tr(\Pi_{n\leq N_B})=\frac{(N_B+1)(N_B+2)}{2}$, the joint state after squashing can be written as
\begin{flalign}
    \widetilde{\rho}_{AB} &=(\mathbb{1}_A\otimes\Lambda)\rho_{AB}\nonumber\\
    &= \sum_{i,j=1}^{d_A}\sum_{n,m=1}^\infty \rho_{i,j}^{n,m} E_{i,j}\otimes \Lambda(E_{n,m})\nonumber\\
    &= \sum_{i,j=1}^{d_A} E_{i,j}\otimes \left(\sum_{n,m=1}^{\Tr(\Pi_{n\leq N_B})} \rho_{i,j}^{n,m} E_{n,m} + \sum_{k=1}^{M_B} c^k_{i,j} \widetilde{E}_{k,k}\right)  \label{SquashedStateMatrixForm}\\
    &= (\mathbb{1}_A\otimes\Pi_{n\leq N_B})\rho_{AB}(\mathbb{1}_A\otimes\Pi_{n\leq N_B})\label{tilderho}\\ 
    &+ \sum_{k=1}^{M_B}\left(\sum_{i=1}^{d_A} c^k_{i,i}  E_{i,i}+ \sum_{i<j}^{d_A} c^k_{i,j} E_{i,j} + (c^k_{i,j})^*  E_{j,i}\right)\otimes\widetilde{E}_{k,k} \nonumber
\end{flalign}
where we define $\widetilde{E}_{k,l} = E_{\Tr(\Pi_{n\leq N_B})+k,\;\Tr(\Pi_{n\leq N_B})+l}$\;, $c^k_{i,j} = \Tr[P_k\;(\sum_{n,m=\Tr(\Pi_{n\leq N_B})+1}^{\infty} \;\rho_{i,j}^{n,m} E_{n,m})]$, and $M_B$ to be the number of POVM elements. Since $\rho_{AB}$ is Hermitian, we also know that 
\begin{equation}
    (\rho_{i,j}^{n,m})^* = \rho_{j,i}^{m,n} \text{ and } (c^k_{i,j})^* = c^k_{j,i}\;.
\end{equation}
Therefore, we only have to optimize over $(d_A\;\Tr(\Pi_{n\leq N_B}))^2+d_A^2\times M_B$ real parameters instead of $[d_A(\Tr(\Pi_{n\leq N_B})+M_B)]^2$ real parameters if we simply take the squashed state as a $d_A(\Tr(\Pi_{n\leq N_B})+M_B)$-dimensional density matrix before imposing any optimisation constraints. By reducing the number of parameters, we observe a significant speedup in the optimisation (for $d_A = 4$ and $M_B = 28$). \vspace{-10pt}

\subsection{Speedup in checking constraints}\label{App:Speedup2}\vspace{-5pt}

In the optimisation problem (\ref{OPT}), to impose each of the constraints require explicit evaluation of the inner product between the updated squashed state $\rho$ and each constraint matrix $\Gamma_\mu$. As the squashed state and all the constraint matrices in (\ref{OPT}) admit a block-diagonal structure, we only need to consider the matrix elements of $\rho$ and $\{\Gamma_\mu\}$ that are contained in these blocks to calculate the inner product. We will show that by defining new optimisation variables of smaller dimensions, the optimisation problem (\ref{OPT}) can be restructured so that each constraint can be checked faster. By doing so, the optimisation problem can be solved quicker.

Let $\Gamma$ be a squashed constraint matrix, which is Hermitian and can be expressed in the squashed basis as
\begin{equation}
    \Gamma = \sum_{i,j=1}^{d_A} E_{i,j}\otimes \left(\sum_{n,m=1}^{\Tr(\Pi_{n\leq N_B})} \Gamma_{i,j}^{n,m} E_{n,m} + \sum_{k,l=1}^{M_B} \Gamma^{k,l}_{i,j} \widetilde{E}_{k,l}\right)
\end{equation}
where $\Gamma_{i,j}^{n,m} \in \mathbb{C}$, and satisfy $(\Gamma_{i,j}^{n,m})^* = \Gamma_{j,i}^{m,n}$ $\forall$ $i,j,n,m$. We can split $\Tr(\Gamma\widetilde{\rho}_{AB})$ into three terms as in
\begin{flalign}
    \Tr(\Gamma\widetilde{\rho}_{AB}) &= \sum_{i,j=1}^{d_A} \left(\sum_{n,m=1}^{\Tr(\Pi_{n\leq N_B})} \Gamma_{i,j}^{n,m}\rho_{j,i}^{m,n}+ \sum_{k=1}^{M_B}\Gamma_{i,j}^{k,k}c_{j,i}^k\right)\nonumber\\
    &= \Tr(\Gamma\rho_{n\leq N_B}) + \langle \vec{\Gamma}_\text{flag}|\vec{c}_\text{diag}\rangle + 2\text{Re}(\langle \vec{\Gamma}_\text{flag}|\vec{c}_\text{off}\rangle), \label{trGammaRho}
\end{flalign}
where we define $\rho_{n\leq N_B} = (\mathbb{1}_A\otimes\Pi_{n\leq N_B})\;\rho_{AB}\;(\mathbb{1}_A\otimes\Pi_{n\leq N_B})$, $|\vec{\Gamma}_\text{flag}\rangle = \sum_{i,j=1}^{d_A}\sum_{k=1}^{M_B}\Gamma_{i,j}^{k,k}|i\rangle\otimes|j\rangle\otimes|k\rangle$, $|\vec{c}_\text{diag}\rangle = \sum_{i=1}^{d_A}\sum_{k=1}^{M_B}c_{i,i}^k|i\rangle\otimes|i\rangle\otimes|k\rangle$ and $|\vec{c}_\text{off}\rangle = \sum_{i<j}^{d_A}\sum_{k=1}^{M_B}c_{i,j}^k|i\rangle\otimes|j\rangle\otimes|k\rangle$. The expression (\ref{trGammaRho}) requires much fewer calculations in tracing the matrix product in the flag-state subspace (i.e. span$\{\widetilde{E}_{k,l}\}$).

Define a function $\mathcal{R}(\sigma) = D(\mathcal{G}(\sigma)||\mathcal{Z}(\mathcal{G}(\sigma)))$ and an operator-valued function $\mathcal{M}$ which maps $\rho_{n\leq N_B}$, $|\vec{c}_\text{diag}\rangle$ and $|\vec{c}_\text{off}\rangle$ to the density matrix $\widetilde{\rho}_{AB}$ of the form in (\ref{tilderho}) where the coefficients can be retrieved from $c_{i,i}^k = \langle i,i,k|\vec{c}_\text{diag}\rangle$ and $c_{i,j}^k = \langle i,j,k|\vec{c}_\text{off}\rangle$ with $|i,j,k\rangle \coloneqq |i\rangle\otimes|j\rangle\otimes|k\rangle$. The convex optimisation problem can be restructured into
\begin{flalign}
    &\text{minimise \hspace{2pt}} \mathcal{R}\left(\mathcal{M}\left(\rho_{n\leq N_B}\;,\;|\vec{c}_\text{diag}\rangle,\;|\vec{c}_\text{off}\rangle\right)\right)\nonumber\\
    &\text{subject to}\nonumber\\
    & \text{\hspace{12pt}} \Tr(\Gamma_\mu\;\rho_{n\leq N_B}) + \langle \vec{\Gamma}_{\mu,\text{flag}}|\vec{c}_\text{diag}\rangle + 2\text{Re}(\langle \vec{\Gamma}_{\mu,\text{flag}}|\vec{c}_\text{off}\rangle) = \gamma_\mu\;,\nonumber\\
    & \text{\hspace{12pt}} \Tr(\widetilde{\Gamma}_\nu\;\rho_{n\leq N_B}) + \langle \vec{\widetilde{\Gamma}}_{\nu,\text{flag}}|\vec{c}_\text{diag}\rangle + 2\text{Re}(\langle \vec{\widetilde{\Gamma}}_{\nu,\text{flag}}|\vec{c}_\text{off}\rangle) \geq \widetilde{\gamma}_\nu\;,\nonumber\\
    & \text{\hspace{12pt}} \mathcal{M}\left(\rho_{n\leq N_B}\;,\;|\vec{c}_\text{diag}\rangle,\;|\vec{c}_\text{off}\rangle\right) \geq 0,
\end{flalign}
where the free variables for the numerical optimisation are $\rho_{n\leq N_B}\in\mathcal{D}(\mathbb{C}^{d_A\Tr(\Pi_{n\leq N_B})})$, $|\vec{c}_\text{diag}\rangle\in\mathbb{R}^{d_A M_B}$, and $|\vec{c}_\text{off}\rangle\in\mathbb{C}^{d_A(d_A-1)M_B/2}$.

Since the equality and inequality constraints (133 constraints in (\ref{OPT})) have to be checked for each run of the optimisation, reducing the time and memory used in matrix multiplications of $\{\Gamma_\mu\}$ (and $\{\widetilde{\Gamma}_\nu\}$) with the squashed state $\widetilde{\rho}_{AB}$ substantially improves the runtime of the whole key rate calculation. \vspace{-10pt}

\subsection{Speedup in evaluating $D(\mathcal{G}(\rho_{AB})||\mathcal{Z}(\mathcal{G}(\rho_{AB})))$}\label{App:RelEntSec}\vspace{-5pt}

Recall the definitions of the $\mathcal{G}$ and $\mathcal{Z}$ maps as stated in Eqns. (\ref{Gmap}) -- (\ref{Zmap}). Using the form of the shared state $\rho_{AB}$ specified in Eqn. (\ref{SquashedStateMatrixForm}) with $i,j\in\{0,1,2,3\}$ and $M_B = 28$, the state $\mathcal{G}(\rho_{AB})$ can be expanded into
\begin{flalign}
    \mathcal{G}(\rho_{AB}) &= \left[(|0\rangle\langle0|_R \otimes E_{0,0}^A)\otimes \sigma_{0,0} \right. \nonumber\\
    &\text{\hspace{6pt}} + (|0\rangle\langle1|_R \otimes E_{0,2}^A)\otimes \sigma_{0,2} \nonumber\\
    &\text{\hspace{6pt}} + (|1\rangle\langle0|_R \otimes E_{2,0}^A)\otimes \sigma_{2,0} \nonumber\\
    &\text{\hspace{6pt}} + \left. (|1\rangle\langle1|_R \otimes E_{2,2}^A)\otimes \sigma_{2,2} \right]\otimes |0\rangle\langle0|_{\widetilde{B}} \nonumber\\
    &+ \left[(|0\rangle\langle0|_R \otimes E_{1,1}^A)\otimes \sigma_{1,1} \right. \nonumber\\
    &\text{\hspace{6pt}} + (|0\rangle\langle1|_R \otimes E_{1,3}^A)\otimes \sigma_{1,3} \nonumber\\
    &\text{\hspace{6pt}} + (|1\rangle\langle0|_R \otimes E_{3,1}^A)\otimes \sigma_{3,1} \nonumber\\
    &\text{\hspace{6pt}} + \left. (|1\rangle\langle1|_R \otimes E_{3,3}^A)\otimes \sigma_{3,3} \right]\otimes |1\rangle\langle1|_{\widetilde{B}}\;, \label{GrhoExplicit}
\end{flalign}
where $\sigma_{i,j} \coloneqq \mathcal{F}_{\alpha(i)}^B\left(\sum_{n,m=1}^{\Tr(\Pi_{n\leq N_B})} \rho_{i,j}^{n,m} E_{n,m}\right)\mathcal{F}_{\alpha(i)}^B + \mathcal{F}_{\alpha(i)}^B\left(\sum_{k=1}^{28} c^k_{i,j} \widetilde{E}_{k,k}\right)\mathcal{F}_{\alpha(i)}^B$ with $\alpha(i) = i\text{ mod }2$. Apply the $\mathcal{Z}$ map to $\mathcal{G}(\rho_{AB})$ will get
\begin{flalign}
    \mathcal{Z}(\mathcal{G}(\rho_{AB})) &= \left[(|0\rangle\langle0|_R \otimes E_{0,0}^A)\otimes \sigma_{0,0} \right. \nonumber\\
    &\text{\hspace{6pt}} + \left. (|1\rangle\langle1|_R \otimes E_{2,2}^A)\otimes \sigma_{2,2} \right]\otimes |0\rangle\langle0|_{\widetilde{B}} \nonumber\\
    &+ \left[(|0\rangle\langle0|_R \otimes E_{1,1}^A)\otimes \sigma_{1,1} \right. \nonumber\\
    &\text{\hspace{6pt}} + \left. (|1\rangle\langle1|_R \otimes E_{3,3}^A)\otimes \sigma_{3,3} \right]\otimes |1\rangle\langle1|_{\widetilde{B}}\;. \label{ZGrhoExplicit}
\end{flalign}

Since Bob's basis announcement partitions $\mathcal{G}(\rho_{AB})$ into 2 orthogonal subspaces with the orthogonal projections and his quantum system $B$ is further partitioned into 2 orthogonal subspaces (i.e. $(n$\hspace{0pt}$\leq$\hspace{0pt}$N_B)$-photon subspace and the flag-state subspace), $\mathcal{G}(\rho_{AB})$ as shown in Eqn. (\ref{GrhoExplicit}) can be broken down into 4 orthogonal subspaces. 

Restricting to the image of map $\mathcal{G}$, matrices $\mathcal{G}(\rho_{AB})$ and $\mathcal{Z}(\mathcal{G}(\rho_{AB}))$ can be simplified to
\begin{equation}
    \mathcal{G}(\rho_{AB}) = \left(
    \begin{matrix}
        \sigma_{0,0} & \sigma_{0,2} & 0 & 0 \\
        \sigma_{2,0} & \sigma_{2,2} & 0 & 0 \\
        0 & 0 & \sigma_{1,1} & \sigma_{1,3} \\
        0 & 0 & \sigma_{1,3} & \sigma_{3,3}
    \end{matrix}
    \right), \label{GrhoMatrix}
\end{equation}
\begin{equation}
    \mathcal{Z}(\mathcal{G}(\rho_{AB})) = \left(
    \begin{matrix}
        \sigma_{0,0} & 0 & 0 & 0 \\
        0 & \sigma_{2,2} & 0 & 0 \\
        0 & 0 & \sigma_{1,1} & 0 \\
        0 & 0 & 0 & \sigma_{3,3}
    \end{matrix}
    \right). \label{ZGrhoMatrix}
\end{equation}

Recall the definition of relative entropy: $D(\rho||\sigma) = \Tr(\rho\log\rho) - \Tr(\rho\log\sigma)$, which is finite if $\text{ker}(\sigma)\subseteq\text{ker}(\rho)$. We can restrict to non-zero subspaces and express the objective function as in Eqn. (\ref{ObjectFunction}) below.
\begin{flalign}
    &\Tr(\mathcal{G}(\rho_{AB})\log\mathcal{G}(\rho_{AB})) \\
    = &\sum_{i=0}^1 \left[\Tr\left(\tau_i^{n\leq N_B}\log\tau_i^{n\leq N_B}\right) + \Tr\left(\tau_i^{\text{flag}}\log\tau_i^{\text{flag}}\right)\right],\nonumber\\
    &\Tr(\mathcal{G}(\rho_{AB})\log\mathcal{Z}(\mathcal{G}(\rho_{AB}))) \\ 
    = &\sum_{i=0}^1 \left[\Tr\left(\tau_i^{n\leq N_B}\log\mathcal{P}(\tau_i^{n\leq N_B})\right) + \Tr\left(\tau_i^{\text{flag}}\log\mathcal{P}(\tau_i^{\text{flag}})\right)\right],\nonumber\\
    &D(\mathcal{G}(\rho_{AB})||\mathcal{Z}(\mathcal{G}(\rho_{AB}))) \label{ObjectFunction}\\
    = &\sum_{i=0}^1 \left[D\left(\tau_i^{n\leq N_B}||\mathcal{P}(\tau_i^{n\leq N_B})\right) + D\left(\tau_i^{\text{flag}}||\mathcal{P}(\tau_i^{\text{flag}})\right)\right],\nonumber\\
    &\tau_i^{\beta} \coloneqq \left(
    \begin{matrix}
        \sigma_{i,i}^{\beta} & \sigma_{i,i+2}^{\beta}\\
        \sigma_{i+2,i}^{\beta} & \sigma_{i+2,i+2}^{\beta}
    \end{matrix}
    \right), \text{\hspace{5pt}} \mathcal{P}(\tau_i^{\beta}) \coloneqq \left(
    \begin{matrix}
        \sigma_{i,i}^{\beta} & 0\\
        0 & \sigma_{i+2,i+2}^{\beta}
    \end{matrix}
    \right) \nonumber
\end{flalign}
with $\beta\in\{n\leq N_B, \text{flag}\}$, where we define the matrices $\sigma_{i,j}^{n\leq N_B} \coloneqq \mathcal{F}_{\alpha(i)}^B\left(\sum_{n,m=1}^{\Tr(\Pi_{n\leq N_B})} \rho_{i,j}^{n,m} E_{n,m}\right)\mathcal{F}_{\alpha(i)}^B$ and $\sigma_{i,j}^{\text{flag}} \coloneqq \mathcal{F}_{\alpha(i)}^B\left(\sum_{k=1}^{28} c^k_{i,j} \widetilde{E}_{k,k}\right)\mathcal{F}_{\alpha(i)}^B$. 

The objective function in (\ref{ObjectFunction}) only requires diagonalisation and the logarithms of the smaller matrices $\tau_i^{\beta}$ and $\mathcal{P}(\tau_i^{\beta})$ for $i\in\{0,1\}$ and $\beta\in\{n\leq N_B, \text{flag}\}$. Therefore, the expression in (\ref{ObjectFunction}) can be computed much quicker than if we directly calculate the relative entropy with the full matrices $\mathcal{G}(\rho_{AB})$ and $\mathcal{Z}(\mathcal{G}(\rho_{AB}))$.

\subsection{Speedup in evaluating the perturbed objective function}\label{App:Perturb}

In the step of linearising the convex optimisation problem, the gradient of the objective function has to be evaluated at the suboptimal point obtained from the first step \cite{AdamNumerics}. As pointed out in Sec.\;3.2 of Ref.\;\cite{AdamNumerics}, the gradient is undefined if the matrix $\mathcal{G}(\rho_{AB})$ is not full rank. Besides, due to the finite numerical precision of a computer, the computed matrix $\mathcal{G}(\rho_{AB})$ may have negative eigenvalues for which the objective function is undefined. In these cases, we perform a perturbation on the matrix $\mathcal{G}(\rho_{AB})$ by applying a depolarising channel which gives the perturbed map $\mathcal{G}_{\epsilon}(\rho_{AB})$, as defined in \cite{AdamNumerics},
\begin{equation}
\begin{split}
    \mathcal{G}_{\epsilon}(\rho_{AB}) &\coloneqq (1-\epsilon)\mathcal{G}(\rho_{AB}) + \frac{\epsilon}{d'}\mathbb{1}_{d'}\\ 
    &= (1-\epsilon)\mathcal{G}(\rho_{AB}) + \frac{\epsilon}{d'}\mathbb{1}|_{\text{Im}(\mathcal{G})} + \frac{\epsilon}{d'}\mathbb{1}|_{\text{ker}(\mathcal{G})}\;,\label{Gepsilon}
\end{split}
\end{equation}
where $\epsilon>0$ is the perturbation parameter and $d'= \text{dim}(\mathcal{G}(\rho_{AB}))$. Applying the $\mathcal{Z}$ map to (\ref{Gepsilon}) results in
\begin{equation}
    \mathcal{Z}(\mathcal{G}_{\epsilon}(\rho_{AB})) = (1-\epsilon)\mathcal{Z}(\mathcal{G}(\rho_{AB})) + \frac{\epsilon}{d'}\mathbb{1}|_{\text{Im}(\mathcal{G})} + \frac{\epsilon}{d'}\mathbb{1}|_{\text{ker}(\mathcal{G})}\;. \label{ZGepsilon}
\end{equation}

The new objective function $D(\mathcal{G}_{\epsilon}(\rho_{AB})||\mathcal{Z}(\mathcal{G}_{\epsilon}(\rho_{AB})))$ is the relative entropy of the two perturbed matrices (\ref{Gepsilon}) and (\ref{ZGepsilon}). We now show that the evaluation of the relative entropy can be restricted to the image of the map $\mathcal{G}$. We evaluate the matrix logarithms
\begin{flalign}
    \log \mathcal{G}_{\epsilon}(\rho_{AB}) &= \log\left[(1-\epsilon)\mathcal{G}(\rho_{AB})+ \frac{\epsilon}{d'}\mathbb{1}|_{\text{Im}(\mathcal{G})}\right] \nonumber\\
    &\;+ \log\left(\frac{\epsilon}{d'}\mathbb{1}|_{\text{ker}(\mathcal{G})}\right), \label{logGepsilon}
\end{flalign}
\begin{flalign}
    \log \mathcal{Z}(\mathcal{G}_{\epsilon}(\rho_{AB})) &= \log\left[(1-\epsilon)\mathcal{Z}(\mathcal{G}(\rho_{AB})) + \frac{\epsilon}{d'}\mathbb{1}|_{\text{Im}(\mathcal{G})}\right]\nonumber\\ 
    &+ \log\left(\frac{\epsilon}{d'}\mathbb{1}|_{\text{ker}(\mathcal{G})}\right).\label{logZGepsilon}
\end{flalign}
and define $\widetilde{\mathcal{G}}_{\epsilon}(\rho_{AB}) \coloneqq \Pi_{\text{Im}(\mathcal{G})}\mathcal{G}_{\epsilon}(\rho_{AB})\Pi_{\text{Im}(\mathcal{G})}$ to obtain
\begin{flalign}
    & \hspace{1pt}D(\mathcal{G}_{\epsilon}(\rho_{AB})||\mathcal{Z}(\mathcal{G}_{\epsilon}(\rho_{AB}))) \nonumber\\ 
    = & \hspace{1pt}\Tr\{\mathcal{G}_{\epsilon}(\rho_{AB})\left[\log \mathcal{G}_{\epsilon}(\rho_{AB}) - \log \mathcal{Z}(\mathcal{G}_{\epsilon}(\rho_{AB}))\right]\} \label{fullGepsilon}\\
    = & \hspace{1pt}\Tr\{\widetilde{\mathcal{G}}_{\epsilon}(\rho_{AB})[\log \widetilde{\mathcal{G}}_{\epsilon}(\rho_{AB}) - \log \mathcal{Z}(\widetilde{\mathcal{G}}_{\epsilon}(\rho_{AB}))]\} \label{subspaceGepsilon}\\
    = & \hspace{1pt}D(\widetilde{\mathcal{G}}_{\epsilon}(\rho_{AB})||\mathcal{Z}(\widetilde{\mathcal{G}}_{\epsilon}(\rho_{AB}))). \label{RelEntTildeG}
\end{flalign}
The step going from (\ref{fullGepsilon}) to (\ref{subspaceGepsilon}) comes from the fact that Eqn. (\ref{logGepsilon}) minus (\ref{logZGepsilon}) results in the zero operator in the kernel of map $\mathcal{G}$. Now that we only have to consider the image of $\mathcal{G}$ in (\ref{RelEntTildeG}), we can use the decomposition described in Eqn. (\ref{ObjectFunction}) but with the matrices $\tau_i^{\beta}$ and $\mathcal{P}(\tau_i^{\beta})$ replaced by $\widetilde{\tau}_i^{\beta}$ and $\mathcal{P}(\widetilde{\tau}_i^{\beta})$ respectively, which are defined as
\begin{flalign}
    \widetilde{\tau}_i^{\beta} &\coloneqq (1-\epsilon)\tau_i^{\beta} + \frac{\epsilon}{d'}(\mathbb{1}_{\beta}\oplus\mathbb{1}_{\beta}), \\ 
    \mathcal{P}(\widetilde{\tau}_i^{\beta}) &\coloneqq (1-\epsilon)\mathcal{P}(\tau_i^{\beta}) + \frac{\epsilon}{d'}(\mathbb{1}_{\beta}\oplus\mathbb{1}_{\beta})
\end{flalign}
with $\beta\in\{n\leq N_B, \text{flag}\}$. Since we can break down the evaluation of the perturbed objective function into calculations on restricted subspaces, the speedup described in Appendix \ref{App:RelEntSec} applies here.

\bibliography{unbalBB84cite}

\end{document}